\documentclass[aps,pre,twocolumn,showpacs,superscriptaddress,groupedaddress]{revtex4-1}
\pdfoutput=1
\usepackage{hyperref}
\usepackage{epsfig}
\usepackage{amsmath}
\usepackage{graphicx}
\usepackage{wrapfig}
\usepackage{dcolumn}
\usepackage{bm}
\usepackage{amssymb}
\usepackage{titlesec}
\usepackage{mathtools}
\usepackage{relsize}
\usepackage{cleveref}
\usepackage[usenames,dvipsnames]{color}
\usepackage{ulem}

\begin{document}
\title{Network desynchronization by non-Gaussian fluctuations}
\author{Jason Hindes$^{1}$, Philippe Jacquod$^{2,3}$, and Ira B. Schwartz$^{1}$}
\affiliation{$^{1}$U.S. Naval Research Laboratory, Code 6792, Plasma Physics Division, Washington, DC 20375, USA}
\affiliation{$^{2}$School of Engineering, University of Applied Sciences of Western Switzerland HES-SO, CH-1951 Sion, Switzerland}
\affiliation{$^{3}$Department of Quantum Matter Physics, University of Geneva, CH-1211 Geneva, Switzerland}

\begin{abstract}
Many networks must maintain synchrony despite the fact that they operate in noisy environments. 
Important examples are stochastic inertial oscillators, which are known to
exhibit fluctuations with broad tails in many applications, including electric power networks with renewable energy sources. 
Such non-Gaussian fluctuations can result in rare network
desynchronization. Here we build a general theory for inertial oscillator
network desynchronization by non-Gaussian noise. We compute the rate of desynchronization
and show that higher-moments of noise enter at specific powers of coupling:
either speeding up or slowing down the rate exponentially depending on how
noise statistics match the statistics of a network's slowest mode. Finally, we use our theory to introduce a technique that drastically reduces the effective description of network desynchronization.
Most interestingly, when instability is associated with a single edge, the reduction is to one stochastic oscillator. 



\end{abstract}
\maketitle


\section{\label{sec:Intro}INTRODUCTION}
Networks of coupled oscillators form the basis for complex physical,
biological and technological systems\cite{RN909}, such as smart grids\cite{RN914}, 
Josephson junction arrays\cite{ISI:000258975100054,ISI:A1982NL11700063},
optical networks\cite{RN915}, biological networks\cite{RN910,RN919,RN926,RN925,RN920},
and coupled mechanical devices\cite{RN916,RN918}. From a deterministic viewpoint, most networks operate in stable attractor regimes,
such as synchronized oscillations. 
However, in reality uncertainties and noise produce fluctuations from an attractor, which over long time scales may bring a network into dynamically unstable 
states and result in large deviations. 
Consequently, much recent attention has been given to the effects of noise on networked oscillators, e.g., stochastic escape\cite{Timme2,ISI:000428996900014,Tyloo,HindesKM,DeVille}, noise cancellation\cite{Ronellenfitsch}, noise propagation\cite{ISI:000351565600008,Haehne,Zhang}, and synchronization\cite{ISI:000383052300003,ISI:000417337800032}.

A motivating application for stochastic networked oscillators is that of fluctuating power grids driven by renewable energy sources, such as wind and solar. Such fluctuations are highly non-Gaussian\cite{Milan,Anvari,Schafer}, and may significantly impact power-grid stability\cite{Kamps,RN935,Kurths2,Nesti}. Non-Gaussian noise is understood to exponentially alter the rates for large, rare fluctuations in simple oscillator systems, including Josephson junctions and micro-mechanical oscillators\cite{RN937,RN939,RN940,RN936,RN938,RN941}. Yet, predicting escape from synchrony in complex oscillator networks, subjected to general noise patterns, remains an outstanding problem. 

This work takes a first step in this direction by analyzing desynchronization in networks of inertial oscillators driven by broadly distributed Poisson noise. Our approach is the first to connect desynchronization events to general noise statistics and network modes and subgraphs. We explicitly show how higher fluctuation moments control desynchronization rates near bifurcation points, both speeding up and slowing down rates depending on whether noise and network statistics are aligned. Our analytical methods allow us to predict desynchronization rates from power-fluctuation data, and explain how general noise tends to effectively desynchronize only certain network subgraphs. 
 
Consider a model for $N$ coupled phase oscillators with inertia. We assume that the acceleration of the $i$th oscillator's phase, $\phi_{i}$, is determined by the velocity $v_{i}\equiv\dot{\phi_{i}}$, natural frequencies (input-power) $P_{i}$, and coupling between oscillators $\sum_{j}\!K_{ij}\sin(\phi_{j}-\phi_{i})$. The oscillator dynamics satisfy the second-order differential equation,
\begin{align}
M\dot{v}_{i}+\gamma v_{i}=P_{i}+\sum_{j}K_{ij}\sin(\phi_{j}-\phi_{i}), 
\label{eq:Swing}
\end{align}
where $M$ and $\gamma$ are inertial and damping constants\cite{FN5}. Eq. (\ref{eq:Swing}) gives an approximation to the swing equations describing the transient dynamics of high-voltage electric power grids \cite{Nishikawa,Machowski}, which we discuss as an illustration. In what follows we take $K_{ij}=KA_{ij}$, where $A_{ij}$ is a symmetric adjacency matrix\cite{Timme2}.

In order to study the effects of non-Gaussian fluctuations on Eq.(\ref{eq:Swing}), we take the input power to be
\begin{align}
\label{eq:Power}
P_{i}(t)=\bar{P}_{i}+p_{i}(t), 
\end{align}
where the fluctuation $p_{i}(t)$ has over-damped dynamics
\begin{align}
\label{eq:Pulse}
\dot{p_{i}}=-\alpha p_{i} + \xi_{i}(t), 
\end{align}
with a damping rate $\alpha$ and stochastic drive $\xi_{i}(t)$\cite{Ronellenfitsch,Timme2,Jacquod1,HindesKM,Tyloo}. Without loss of generality, we take the average natural frequency to vanish, $\sum_i\!\bar{P}_i\!=\!0$. For electric power grids this reflects a balance between production and consumption.
 
\begin{figure}[h]
\includegraphics[scale=0.265]{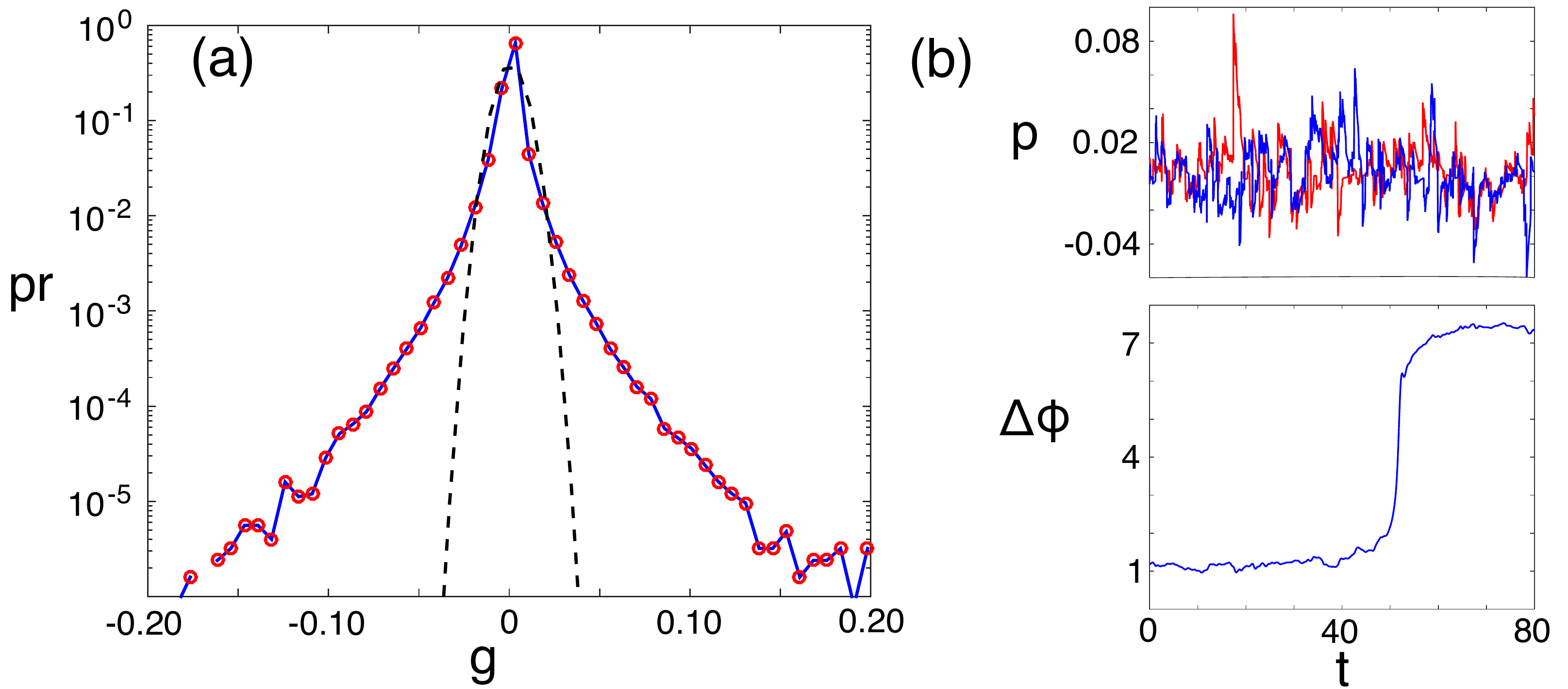}
\caption{Non-Gaussian power fluctuations. (a) Wind turbine power-increment distribution (solid-blue)\cite{Haehne}; Gaussian distribution with the same variance (dashed-black). (b) Fluctuation time series (seconds) given independent and identically distributed Poisson fluctuations for a 30-node network. Amplitudes and rates for Eq.(\ref{eq:XI}) are taken from (a). Power fluctuations for two oscillators are shown in blue and red (top). Phase-difference, $\Delta\phi$, between the oscillators (lower)\cite{FNtime}.}
\label{fig:Noise}
\end{figure}

In modeling the noisy drive $\xi_{i}(t)$ we are interested in cases where $p_{i}(t)$ exhibits large intermittent fluctuations, as seen in the output from wind and solar sources\cite{Kurths2,Milan,Anvari,Schafer}. The non-Gaussian features of such sources can be captured by tracking the change in $p_{i}(t)$ (or power increments) over fixed time-intervals, $\tau$ \cite{Anvari}. In practice, $\tau$ may be set by the time resolution of data. For example, in Fig.\ref{fig:Noise}(a) we show a histogram of power increments, $g(\tau)\!\equiv\!p(t+\tau)-p(t)$, from 12 wind turbines in northern Germany as in\cite{Haehne}. The probability distribution, $\text{pr}(g)$, is highly non-Gaussian, e.g., the fourth moment is much larger than for a Gaussian with the same variance. 

In order to build a flexible noise model that approximates the measured increment distribution, we assign an independent Poisson pulse for every bin, $b$, in the histogram. Each Poisson pulse has an amplitude, $g_{b}$, which is equal to the bin average, and an occurrence rate $\nu_{b}\!=\!\text{pr}(g_{b})/\tau$. Given this choice, a pulse occurs on average every $\tau$ units of time. More generally, let there be $\mathcal{M}$ arbitrary power increments such that the amplitude for the $b$th increment on the $i$th oscillator is $g_{ib}$, where $b\!\in\!\{1,2,...,\mathcal{M}\}$. By denoting the time at which the $n$th such increment occurs as $t_{ib}[n]$, the stochastic drive $\xi_{i}(t)$ can be represented by a sum of Dirac delta functions\cite{Feynman} 
\begin{align}
\label{eq:XI}
\xi_{i}(t)=\sum_{bn}g_{ib}\delta(t-t_{ib}[n]). 
\end{align}
Because the noise $\xi_{i}(t)$ is built from Poisson pulses, each $t_{ib}[n+1]-t_{ib}[n]$ is a stochastic variable with an exponential distribution whose rate is $\nu_{ib}$. For simplicity, we take the time average of the power fluctuations for each oscillator to be zero, i.e., $\sum_{b}g_{ib}\nu_{ib}\!=\!0$. An example time series given our model is shown in Fig.\ref{fig:Noise}(b). 




\section{LARGE FLUCTUATION PICTURE OF DESYNCHRONIZATION}
When the coupling constant $K$ is sufficiently large a synchronized state is a stable
fixed point of Eq.(\ref{eq:Swing}). This stable phase-locked state (PLS) depends on network topology and the
distribution of $\bar{P}_{i}$ \cite{FN5}. In general, the PLS emerges through a saddle-node bifurcation as
$K$ is increased\cite{Mirollo,Manik}, implying the existence of unstable, saddle phase-locked
states. Noise on Eq.(\ref{eq:Swing}) can cause networks to fluctuate to these saddles
\cite{HindesKM,Timme2,Tyloo}. Once a saddle is reached the network can desynchronize by either undergoing a large phase slip upon returning to the PLS modulo $2\pi$, or exiting its basin of attraction altogether. An example is shown in Fig.\ref{fig:Noise}(b)(lower). Such noise-induced desynchronization whereby fluctuations drive oscillator networks to saddle points, are examples of the general phenomenon of basin-escape\cite{Kramers,Mark1}. 


Given the non-Gaussian noise discussed in Sec.\ref{sec:Intro}, our strategy is to construct the
most-likely (optimal) path of noise and network dynamics that maximizes the
probability of reaching a saddle. When desynchronization is rare, the optimal path is
describable using analytical mechanics tools\cite{Kramers,Mark1,WoillezPRL}. Our approach is a generalization of Kramer's theory for escape, and is valid as long as typical fluctuations are small compared to the distances to saddles\cite{Kramers}. 

We begin our analysis with the network probability distribution, $\rho(\boldsymbol{\phi},\bold{v},\bold{p},t)$. It's dynamics satisfy a generalized master equation
\begin{align}
\label{eq:FokkerPlanck} 
&\frac{\partial \rho}{\partial t}= \sum_{i}\!\Bigg[\!-\frac{\partial}{\partial{\phi_{i}}}\!\Big[v_{i}\rho\Big]+\frac{\partial}{\partial{p_{i}}}\!\Big[\alpha p_{i}\rho\Big] \\ \nonumber
&-\frac{\partial}{\partial{v_{i}}}\!\Big[\Big(\!\!-\frac{\gamma v_{i}}{M}+\frac{1}{M}\big(P_{i}+\sum_{j}K_{ij}\sin(\phi_{j}-\phi_{i})\big)\!\Big)\rho\Big] \\ \nonumber  
&+\sum_{b}\nu_{ib}\Big[\rho(\boldsymbol{\phi},\bold{v},\bold{p}-g_{ib}\bold{1}_{i})-\rho(\boldsymbol{\phi},\bold{v},\bold{p})\Big]\!\Bigg], 
\end{align}
where the vector $\bold{1}_{i}\!=\!\left<0{}\;_{1},0{}\;_{2},..,1{}\;_{i},..,0{}\;_{N}\right>$ \cite{LoraPRL2010}. Note that Eq.(\ref{eq:FokkerPlanck}) is similar to a Fokker-Planck equation, except that, instead of the typical diffusive term, there is a sum over discrete increments to $\bold{p}$, given in the last line of Eq.(\ref{eq:FokkerPlanck}), as in a master equation for Poisson processes.   

To analyze rare events encoded in the exponential tail of $\rho(\boldsymbol{\phi},\bold{v},\bold{p},t)$, we substitute a WKB ansatz, $\rho(\boldsymbol{\phi},\bold{v},\bold{p},t)\!\cong\!B\exp\{-S(\boldsymbol{\phi},\bold{v},\bold{p},t)\}$, into Eq.(\ref{eq:FokkerPlanck}), assuming $S(\boldsymbol{\phi},\bold{v},\bold{p},t)\gg1$, and keep the leading order terms in $\partial_{\boldsymbol\phi}S$, $\partial_{\boldsymbol v}S$, and $\partial_{\boldsymbol p}S$ \cite{LoraPRL2010,Mark2}. This approximation converts Eq.(\ref{eq:FokkerPlanck}) to a Hamilton-Jacobi equation (HJE) for the probability exponent, $S(\boldsymbol{\phi},\bold{v},\bold{p},t)$, called the {\it action}, in terms of $\boldsymbol{\phi}$, $\bold{v}$, $\bold{p}$, and their conjugate momenta: $\boldsymbol{\lambda}^{\phi}\equiv\partial_{\boldsymbol\phi}S$, $\boldsymbol{\lambda}^{v}\equiv\partial_{\boldsymbol v}S$, and $\boldsymbol{\lambda}^{p}\equiv\partial_{\boldsymbol p}S$.
The network Hamiltonian is
\begin{align}
&H(\boldsymbol{\phi},\bold{v},\bold{p},\boldsymbol{\lambda}^{\phi},\boldsymbol{\lambda}^{v},\boldsymbol{\lambda}^{p})=\sum_{i}\!\!\Bigg[\;\lambda^{\phi}_{i}v_{i}-\alpha p_{i}\lambda^{p}_{i} \nonumber \\ 
&+\;\sum_{b}\nu_{ib}\Big(\!\exp\{g_{ib}\lambda_{i}^{p}\}-1\Big)\;\;+\nonumber \\
&\frac{\lambda_{i}^{v}}{M}\Big(\!\! -\gamma v_{i} + \bar{P}_{i}+p_{i}+\!\sum_{j}K_{ij}\sin(\phi_{j}-\phi_{i}) \!\Big)\Bigg].
\label{eq:Hamiltonian}  
\end{align}

For the given function $H(\boldsymbol{\phi},\bold{v},\bold{p},\boldsymbol{\lambda}^{\phi},\boldsymbol{\lambda}^{v},\boldsymbol{\lambda}^{p})$,
the optimal-path dynamics satisfy Hamilton's equations\cite{FN2}. Once Hamilton's equations are solved, the action can be calculated as   
\begin{align}
\label{eq:Action}
 S(\boldsymbol{\phi},\bold{v},\bold{p})=\sum_{i}\!\Bigg[\!\int\!\lambda_{i}^{\phi}d\phi_{i}\;+\!\int\!\lambda_{i}^{v}dv_{i}\;+\!\int\!\lambda_{i}^{p}dp_{i}\Bigg]. 
\end{align}
In general, solutions are computable numerically subject to boundary conditions\cite{LindleyIAMM}. Once the action is known, so is the expected waiting time (or inverse rate) for desynchronization\cite{Kramers,Mark1,Mark,HindesKM} 
\begin{align}
\ln{\!\left<T\right>}\approx S(\boldsymbol{\phi}^{s},\bold{0},\bold{0})+\text{constant}.
\end{align}

\subsection{Optimal paths near bifurcation}
Using our desynchronization mechanics, let us first consider the optimal path
(OP) from PLS, $\boldsymbol{\phi}^{*}$, to saddles, $\boldsymbol{\phi}^{s}$,
near the saddle-node bifurcation (SN). We denote the critical coupling
$K_{\text{SN}}$, where $K\!=\!K_{\text{SN}}[1+\kappa]$. When $\kappa\!\ll\!1$,
the dynamics slows onto a one-dimensional manifold with universal
properties. As we will show, statistical moments of the noise 
first contribute to the action at specific powers of $\kappa$, from which we
can calculate their effects on desynchronization rates. In order to simplify the analysis, we assume that the noise for all nodes is independent and identically distributed. Hence, we drop the subscript $i$ in $\nu$, $g$ and $\mu$ from now on.    

First, we construct the lowest-order solution and thereby demonstrate our natural expansion in $\kappa^{1/2}$. Higher order terms and further calculation details are given in Apps.\ref{sec:SNPaths}-\ref{sec:NGE}. From the fixed-point boundary conditions we expand $\boldsymbol{\phi}^{*}$ and $\boldsymbol{\phi}^{s}$ around the SN value, $\boldsymbol{\phi}^{SN}$, in powers of $\kappa$. At the saddle-node bifurcation, the Fiedler mode\cite{Fiedler} of the network Laplacian, $L_{ij}(\boldsymbol{\phi}^{*})\!=A_{ij}\cos(\phi_{j}^{*}-\phi_{i}^{*})-\delta_{ij}\sum_{k}A_{ik}\cos(\phi_{k}^{*}-\phi_{i}^{*})$, has zero eigenvalue. In general, the Fiedler mode is the slowest mode of $L_{ij}(\boldsymbol{\phi}^{*})$, and we denote its components $r_{i}$. The normalized Fiedler mode is defined up to an overall sign, and we choose the convention $r_{i}\!>\!0$ if $\phi_{i}^{s}\!-\!\phi_{i}^{*}\!>\!0$, so that $\bold{r}$ points from $\boldsymbol{\phi}^{*}$ to $\boldsymbol{\phi}^{s}$. Quite remarkably, the fixed points close to bifurcation can be expressed in terms of $r_{i}$, $\phi_{i}^{*}\!=\!\phi_{i}^{SN}\!\!-\!C\kappa^{1/2}r_{i}$ and $\phi_{i}^{s}\!=\!\phi_{i}^{SN}\!\!+\!C\kappa^{1/2}r_{i}$, where:
\begin{align}
\label{eq:C}
 C=\sqrt{\frac{2\big|\sum_{ij}A_{ij}\sin(\phi_{j}^{SN}\!\!-\!\phi_{i}^{SN})r_{i}\big|}{\big|\sum_{ij}A_{ij}\sin(\phi_{j}^{SN}\!\!-\!\phi_{i}^{SN}){[r_{j}\!-r_{i}]}^{2}r_{i}\big|}}.
\end{align}
Because the sums in Eq.(\ref{eq:C}) appear when considering the noise moments, we write $C\!\equiv\!\sqrt{2R_{0}/R_{2}}$. Next, we write the phases emerging from the SN in terms of a coordinate $x(t)$, $\phi_{i}(t)\!=\!\phi_{i}^{SN}\!+\!C\kappa^{1/2}r_{i}x(t)$, where $x\!\in\![-1,1]$, which is valid near bifurcation\cite{Kuznetsov1}. Substituting this form into Hamilton's equations, and collecting terms at order $\kappa$, we find: 
\begin{align}
\label{eq:OP_SN}
\frac{v_{i}}{r_{i}}&=\frac{K_{\text{SN}}\kappa R_{0}[1\!-\!x^{2}]}{\gamma}, \;\;\;\;\;\;\;\;\;\frac{p_{i}}{r_{i}}=2K_{\text{SN}}\kappa R_{0}[1\!-\!x^{2}], \nonumber \\
\frac{\lambda_{i}^{\phi}}{r_{i}}&=\!\frac{2K_{\text{SN}}\kappa R_{0}\gamma\alpha^{2}[1\!-\!x^{2}]}{\mu_{2}}, \;\frac{\lambda_{i}^{v}}{r_{i}}=\!\frac{2K_{\text{SN}}\kappa R_{0}M\alpha^{2}[1\!-\!x^{2}]}{\mu_{2}}\!, \nonumber \\
\frac{\lambda_{i}^{p}}{r_{i}}&=\!\frac{2K_{\text{SN}}\kappa R_{0}\alpha[1\!-\!x^{2}]}{\mu_{2}}, \;\; \dot{x}=\frac{K_{\text{SN}}\kappa^{1/2}\sqrt{2R_{0}R_{2}}[1\!-\!x^{2}]}{2\gamma}\!,
\end{align}
\noindent where $\mu_{2}\!\equiv\!\sum_{b}\nu_{b}g_{b}^{2}$ is the noise variance. From Eqs.(\ref{eq:Action}) and (\ref{eq:OP_SN}), we get 
\begin{equation}
\label{eq:ActionSN}
S(\boldsymbol{\phi}^{s},\bold{0},\bold{0})\approx\frac{8\sqrt{2}K_{\text{SN}}\kappa^{3/2}\gamma\alpha^{2}R_{0}^{3/2}}{3R_{2}^{1/2}\mu_{2}}. 
\end{equation}

The structure of Eq.(\ref{eq:ActionSN}) is interesting. The action at lowest order in $\kappa$ is proportional to the damping rate for power-fluctuations squared, implying that doubling the rate, increases the time-scale for desynchronization as $\left<T\right>\!\rightarrow\!\left<T\right>^{4}$. What's more, each node's contribution is proportional to $\mu_{2}^{\;-1}$, and hence noise distributions with the same variance produce the same rate of desynchronization at lowest order. 
Therefore, the effects of higher fluctuation moments must appear at (and be suppressed by) higher powers of $\kappa$.
 
Continuing the OP construction at higher-powers in $\kappa$, we notice that the lowest-order contribution to the action from the $n$th moment of the noise, $\mu_{n}\!\equiv\!\sum_{b}\nu_{b}g_{b}^{n}$, is determined by energy-conservation for large fluctuations, $H(\boldsymbol{\phi},\bold{v},\bold{p},\boldsymbol{\lambda}^{\phi},\boldsymbol{\lambda}^{v},\boldsymbol{\lambda}^{p})\!=\!0$, at $\mathcal{O}(\kappa^{n})$ \cite{FN2}. Interestingly, such contributions, $\Delta^{(n)}S$, only depend on Eq.(\ref{eq:OP_SN}) and can be calculated (see App.\ref{sec:NGE} for derivation):  
\begin{align}
\label{eq:Correction}
\Delta^{(n)}S=&-\frac{\mu_{n}[\sum_{i}{r_i}^{n}]}{\mu_{2}^{n}}\;\cdot\;\kappa^{n-\frac{1}{2}}K_{\text{SN}}^{n-1}\;\cdot \\
&\frac{2^{n+\frac{1}{2}}\gamma \alpha^{n} R_{0}^{n-\frac{1}{2}}\int_{-1}^{1}(1-x^{2})^{n-1}dx}{R_{2}^{1/2}n!}. \nonumber
\end{align}
\begin{figure}[t]
\includegraphics[scale=0.2215]{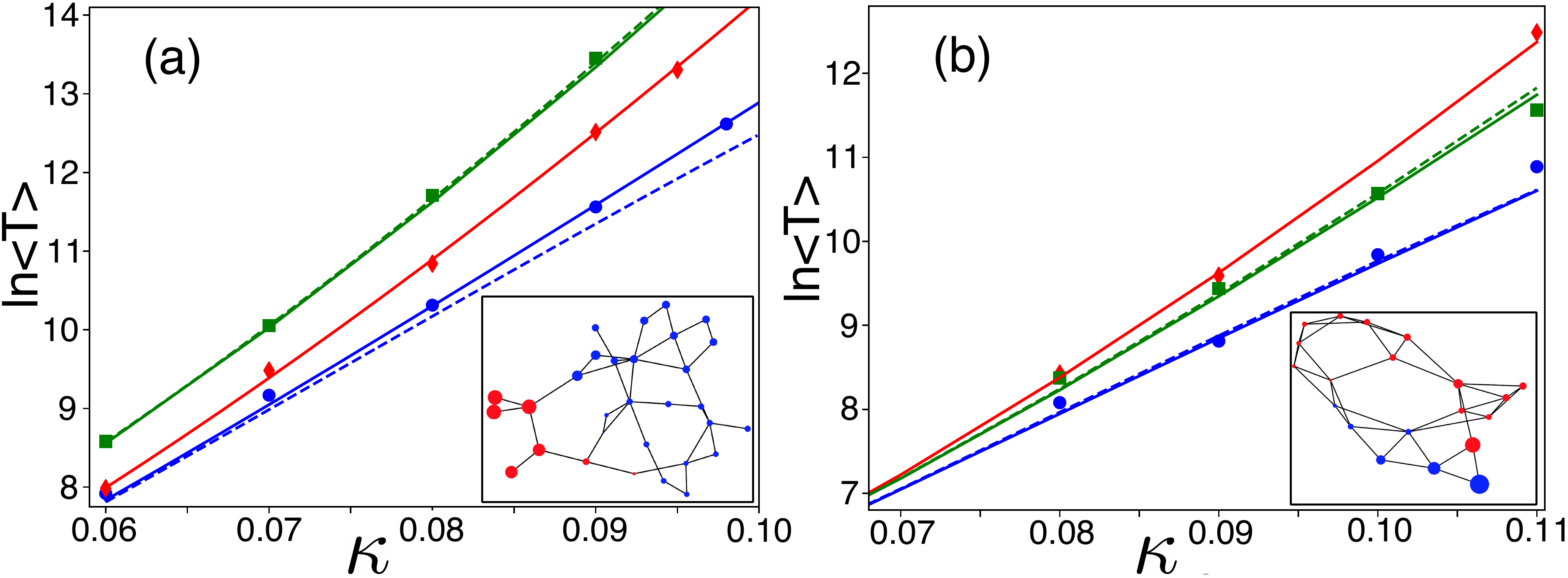}
\caption{Effect of non-Gaussian noise on desynchronization times. Points denote Monte-Carlo simulations and solid lines denote least-action computations, Eq.(\ref{eq:Action}). Fluctuations follow the color scheme: Gaussian (red), positively skewed (green), and symmetric wind turbine (blue)-- all with the same variance. Dashed lines are predictions from Eq.(\ref{eq:Correction}). (a) network with a negatively skewed Fiedler mode. (b) network with a positively skewed Fiedler mode.}
\label{fig:StatMatch}
\end{figure}  

Equation (\ref{eq:Correction}) is very useful for comparing non-Gaussian and Gaussian white noise (GWN) effects. In the latter, the stochastic drive in Eq.(\ref{eq:Pulse}) is replaced by a Gaussian process with time-correlation $\left<\xi_{i}(t)\xi_{j}(t')\right>\!=\!\mu_{2}\delta_{ij}\delta(t-t')$ $\forall\{i,j\}$, and Eq.(\ref{eq:FokkerPlanck}) becomes a Fokker-Planck equation. Simple GWN is considered in most works on stochastic oscillators\cite{Timme2,Tyloo,ISI:000383052300003}. Because the action for GWN is equivalent to keeping only the variance in the Hamiltonian, Eq.(\ref{eq:Correction}) represents the correction to the GWN action from fluctuations with a first non-zero moment $\mu_{n}$ ($n\!>\!2$). 

Interestingly, we see that desynchronization rates exhibit a critical behavior characterized by a spectrum of exponents $n-\frac{1}{2}$ depending on the $n$th moment of the noise distribution. Most importantly, the sign of the first non-Gaussian correction in Eq.(\ref{eq:Correction}) is given by the product of the $n$th noise moment times the $n$th moment of the Fiedler components. In the particular case when $n\!=\!3$, the sign of $\Delta S$ is determined by the product of the skewness of the Fiedler components with that of the noise. When the two skewnesses are aligned, $\Delta S^{(3)} \!<\!0$, and desynchronization occurs at an exponentially faster rate. When they have opposite signs, $\Delta S^{(3)} \!>\!0$, and desynchronization occurs at an exponentially slower rate. Generally speaking, the Fiedler modes in high voltage power grids mostly reside on peripheral nodes - but not all peripheral nodes\cite{PhilippePlosOne}. Hence, such networks are highly skewed, by our definition, and will show exponential sensitivity in desynchronization rates to noise skewness.   

Two examples are shown in Fig.\ref{fig:StatMatch}, where we compare desynchronization times for GWN (red), positively skewed $\mu_{3}\!>\!0$ (green), and wind turbine (blue) fluctuations. All power fluctuations have the same variance. The skewed noise distribution is a simple two-pulse model. Solid lines denote Eq.(\ref{eq:Action}) computations, while points indicate Monte-Carlo simulations. The dashed lines in green and blue denote the action for GWN plus Eq.(\ref{eq:Correction}), which is in good agreement (in most cases hard to distinguish). The networks are drawn in each subpanel. Nodes are blue if $r_{i}\!>\!0$ and red if $r_{i}\!<\!0$; sizes are proportional to $|r_{i}|$. In Fig.\ref{fig:StatMatch}(a), the network has a {\it negatively skewed Fiedler mode}, $\sum_{i}\!{r_i}^{3}\!<\!0$, and hence desynchronization occurs at an exponentially slower rate, just as predicted. In contrast in Fig.\ref{fig:StatMatch}(b) the network has a positively skewed Fiedler mode, and therefore desynchronization occurs at an exponentially faster rate. Note that for both networks, blue lines are always under the red, meaning symmetric noise ($\mu_{3}\!\approx\!0$) produces an increase in desynchronization rates.

\section{SYNCHRONIZED SUBGRAPH APPROXIMATION}
Since we have shown that rare desynchronization occurs along the Fiedler mode near bifurcation, we expect that nodes that are topologically nearby in the network and have similar Fiedler-mode values, do not tend to desynchronize during a large fluctuation -- as a first approximation for general coupling strengths. Partitioning the network according to the Fiedler mode\cite{Fiedler} at bifurcation is a useful approximation for finding much lower-dimensional desynchronization pathways, which we call the synchronized subgraph approximation (SSA). An algorithm for constructing Hamilton's equations for a SSA is given in the App.\ref{sec:SSA}.   


\begin{figure}[t]
\includegraphics[scale=0.2665]{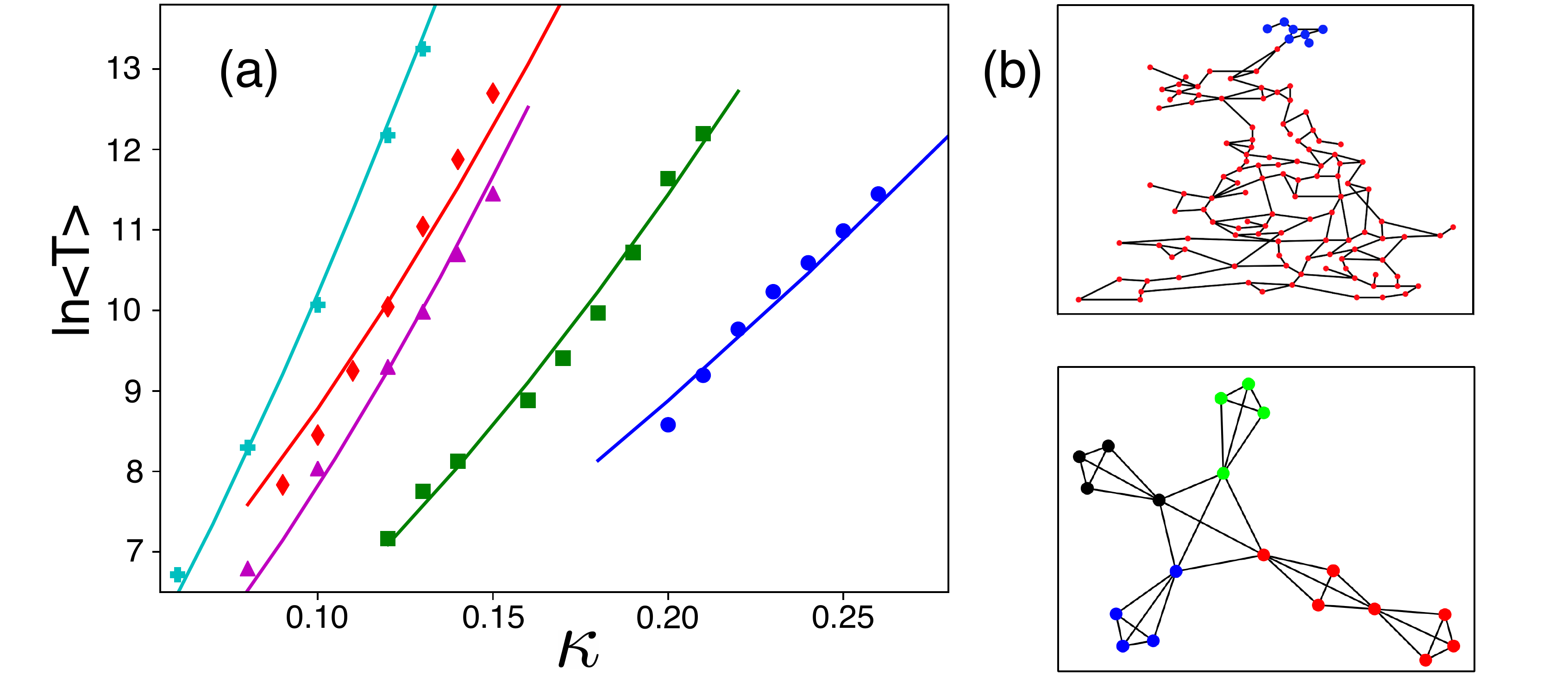}
\caption{Desynchronization times predicted assuming synchronized subgraphs; (a) UK grid with one overloaded edge (green), a 50-node tree (red), a 100-node tree (blue), a block network (cyan), and the UK grid without symmetry at bifurcation (magenta). Solid lines denote computations, while points correspond to Monte-Carlo. (b) Synchronized subgraphs for the UK grid (top), and the block network (bottom). Subgraphs are drawn with different colors.}
\label{fig:SSA}
\end{figure} 

First, the SSA can be made exact for networks where the SN corresponds to a single (overloaded) edge with a phase-difference $\pi/2$. This condition is satisfied for tree topologies and frequently satisfied for sparse networks\cite{Manik}. For such {\it single-cut saddle-nodes} (SCSN), we can always construct a SSA where the network splits into exactly {\it two subgraphs} at bifurcation and the nodes within each remain synchronized on average. If we denote the two subgraphs divided by the overloaded edge $\mathcal{S}_{1}$ and $\mathcal{S}_{2}$, then we can reduce Hamilton's equations to a single noisy oscillator system in relative phase-space coordinates:
\begin{align}
\label{eq:SCSNeff}
\dot{V}&=\frac{1}{M}\Big[-\gamma V + P +\frac{K_{\text{SN}}N}{|\mathcal{S}_{1}||\mathcal{S}_{2}|}\big(1-(1+\kappa)\cos\{\Psi\}\big)\Big] \nonumber \\
\dot{P}&=-\alpha P+\sum_{m}\nu_{m}g_{m}\!\Big[e^{g_{m}\Lambda_{P}/|\mathcal{S}_{1}|}\!-e^{-g_{m}\Lambda_{P}/|\mathcal{S}_{2}|}\Big] \nonumber \\
\dot{\Lambda}_{\Psi}&=-\frac{K_{\text{SN}}(1+\kappa)}{M}\frac{N}{|\mathcal{S}_{1}||\mathcal{S}_{2}|}\Lambda_{V}\sin\{\Psi\}, \;\;\;\; V=\dot{\Psi} \nonumber \\
\dot{\Lambda}_{V}&=-\Lambda_{\Psi}+\frac{\gamma}{M}\Lambda_{V}, \;\;\;\;\;\;\;\;\;\;\;  \dot{\Lambda}_{P}=\alpha\Lambda_{P}-\frac{1}{M}\Lambda_{V}
\end{align}
In Eqs.(\ref{eq:SCSNeff}) $\Psi$ is the phase difference between the two subgraphs, $V$ is the difference in their velocities, $P$ is the difference in the power fluctuations, and the $\Lambda$'s are conjugate momenta.   
The system is closed since $K_{\text{SN}}\!=\!|\sum_{i\in\mathcal{S}_{1}}\bar{P}_{i}|$ for SCSN \cite{Dekker}. Remarkably, we can see that the dynamics is parameterized only by $\kappa$ and the sizes of the two subgraphs, $|\mathcal{S}_{1}|$ and $|\mathcal{S}_{2}|$, but is otherwise {\it independent of topology}. The action corresponds to a single stochastic oscillator:  
\begin{equation}
\label{eq:ActionSCSN}
S(\Psi,V,P)=\int\!\Lambda_{\Psi}d\Psi +\!\int\!\Lambda_{V}dV\!+\! \int\!\Lambda_{P}dP.
\end{equation}

Comparisons between desynchronization times and SSA predictions are shown in
Fig.\ref{fig:SSA}(a) for three networks with a SCSN: red, blue, and green. In
each case $N\!\sim\!\mathcal{O}(100)$. In particular, the UK power grid is
shown in green, and its two subgraphs are drawn in the top of
Fig.\ref{fig:SSA}(b); the red and blue examples in Fig.\ref{fig:SSA}(a) correspond to two randomly generated 100-node trees with power-law degree distributions, and an exponent equal to three. Theory computations are shown with solid lines
(Eq.\ref{eq:ActionSCSN}) and Monte-Carlo simulations with points. Examples show quantitative agreement despite two-orders of magnitude reduction in the effective dynamics. Moreover, in App.\ref{sec:Hetero} we show that Eqs. (\ref{eq:Correction}) and (\ref{eq:ActionSCSN}) are accurate even in networks with significant variation in $M$, $\gamma$, and $\alpha$.  

Similarly, the SSA can be applied to networks without a SCSN, but where there is an exact symmetry in the Fiedler mode at bifurcation. An example is shown in Fig.\ref{fig:SSA}(a) in cyan for a block network. In this case the network splits into exactly four subgraphs at bifurcation. The Fiedler-mode has only four unique component values-- one for each subgraph. The subgraphs are drawn in the lower panel of Fig.\ref{fig:SSA}(b). As a consequence, the SSA consists of four effectively coupled oscillators. 

Lastly, the SSA can be applied to networks whose Fiedler modes do not have exact symmetry at bifurcation, by partitioning the network into subgraphs with approximately uniform Fiedler-mode values. An example is shown in Fig.\ref{fig:SSA}(a) (magenta), again for the UK power grid, but with a different distribution of $\bar{P}_{i}$. In this case, there is no symmetry at bifurcation and each node has a unique Fiedler component value. The solid line represents a SSA assuming 20 subgraphs.    
 
\section{CONCLUSION}
In this work, we analyzed desynchronization in complex
  oscillator networks by non-Gaussian noise. It is often thought that
  broad-tailed fluctuations perturb dynamical systems more strongly. However,
  we found that such noise does not always increase the rate of network
  desynchronization; the latter depends on whether higher-moments of
  fluctuations and a network's slowest mode have the same sign, and is
  therefore a topological as well as a noise effect. Our approach was based on
  arbitrarily distributed Poisson fluctuations, which we fit to power-increment data from renewable energy sources. In addition, we developed a reduction technique for predicting desynchronization, based on the Fiedler mode at bifurcation, and the observation that noise tends to effectively desynchronize only certain network subgraphs. Such a reduction should be valuable for studying rare processes in high-dimensional oscillator networks more broadly, where predicting rare events is both analytically and computationally difficult. 


Lastly, our approach revealed a spectrum of scaling exponents that determine at what powers in the coupling the $n$th moment of noise contributes to desynchronization rates. Our results are general for escape through a saddle. However, our methods can be further generalized to rare events induced by non-Gaussian noise in other dynamical processes in networks including: extinction\cite{HindesPRL2016,HindesPRL2019}, switching\cite{Havlin2017}, and more general oscillator transitions\cite{Loos,HindesChaos}. 


JH was supported through the U.S Naval Research Laboratory Karle Fellowship. PJ was supported by the Swiss National Science Foundation grants (200020\_182050) and (PYAPP2\_154275). IBS was supported by the U.S. Naval Research Laboratory funding (N0001419WX00055) and the Office of Naval Research (N0001419WX01166) and (N0001419WX01322).\\

\appendix{\centerline{\bf V. APPENDIX}}
\section{\label{sec:OPs}Computing optimal paths}
Optimal paths for rare desynchronization satisfy Hamilton's equations:  
\begin{align}
\label{eq:FullHamiltons1}
\dot{\phi}_{i}=&\;v_{i}, \\
\label{eq:FullHamiltons2}
M\dot{v}_{i}=&-\gamma v_{i}+p_{i}+\bar{P}_{i}+K\sum_{j}A_{ij}\sin(\phi_{j}-\phi_{i}), \\
\label{eq:FullHamiltons3}
\dot{p}_{i}=&-\alpha p_{i} +\sum_{b}\nu_{ib}g_{ib}\exp\{g_{ib}\lambda_{i}^{p}\},\\
\label{eq:FullHamiltons4}
\dot{\lambda}_{i}^{\phi}=&-\frac{K}{M}\sum_{j}A_{ij}\cos(\phi_{j}-\phi_{i})[\lambda_{j}^{v}-\lambda_{i}^{v}],\\
\label{eq:FullHamiltons5}
\dot{\lambda}_{i}^{v}=&-\lambda_{i}^{\phi} +\frac{\gamma}{M}\lambda_{i}^{v},\\
\label{eq:FullHamiltons6}
\dot{\lambda}_{i}^{p}=&\alpha\lambda_{i}^{p} -\lambda_{i}^{v}/M.
\end{align}

Numerical solutions of Eqs.(\ref{eq:FullHamiltons1}-\ref{eq:FullHamiltons6}) were found using the Iterative-Action-Minimization-Method (B. S. Lindley and I. B. Schwartz, Physica D {\bf255}, 22 (2013)) with fixed-point boundary conditions. Example Matlab code can be found in the supplementary material of (J. Hindes and I. B. Schwartz, EPL {\bf 120}, 56004 (2017)), for instance, and is available upon request. The method requires a trial solution. For small $\kappa$ we used Eqs.(\ref{eq:OP_SN}) from the main text, and then bootstrapped to other regions of parameter space.  

\section{\label{sec:FP} Fixed points}
The coupling at which the saddle-node occurs, $K_{\text{SN}}$, can be computed numerically by solving the following $N+1$ equations with a quasi-Newton method: 
\begin{align}
\label{eq:SN1}
0=&\bar{P}_{i}+K_{\text{SN}}\sum_{j}A_{ij}\sin(\phi_{j}^{SN}-\phi_{i}^{SN}) \;\; \forall i \\
\label{eq:SN2}
0=&z;
\end{align}
where $z$ is the second smallest (in magnitude) eigenvalue of the Laplacian matrix $L_{ij}(\boldsymbol{\phi}^{SN})\!=A_{ij}\cos(\phi_{j}^{SN}-\phi_{i}^{SN})-\delta_{ij}\sum_{k}A_{ik}\cos(\phi_{j}^{SN}-\phi_{i}^{SN})$. 

In order to calculate optimal paths as a function of the distance to bifurcation $\kappa$, where $K\!=\!K_{\text{SN}}[1+\kappa]$, we first calculate the fixed points of Eq.(\ref{eq:Swing}) in powers of $\kappa$. Let us substitute $\phi_{i}^{*}\!=\!\phi_{i}^{SN}\!+\!\kappa^{1/2}q_{i}\!+\!\kappa w_{i}\!+\!...$ into Eq.(\ref{eq:Swing}), given $\dot{v}_{i}\!=\!0$ and $v_{i}\!=\!0$ $\forall i$. Note, the sub(super)-script SN implies evaluation at the saddle-node. Our goal is to find $q_{i}$. At $\mathcal{O}(\kappa^{1/2})$ we find the equation
\begin{align}
\label{eq:FixedPoint0p5}
0_{i}=\sum_{j}A_{ij}\cos(\phi_{j}^{SN}\!-\!\phi_{i}^{SN})[q_{j}-q_{i}],
\end{align}
which expresses the saddle-node condition that the network Laplacian has a Fiedler mode with eigenvalue zero. Hence, we may write $q_{i}\!=\!-Cr_{i}$, where $r_{i}$ is the Fiedler mode at bifurcation and $C$ is a constant that we wish to determine. Continuing to $\mathcal{O}(\kappa)$, gives 
\begin{align}
\label{eq:FixedPoint1p0}
0_{i}=&\sum_{j}A_{ij}\cos(\phi_{j}^{SN}\!-\!\phi_{i}^{SN})[w_{j}-w_{i}] \nonumber \\  
&+\sum_{j}A_{ij}\sin(\phi_{j}^{SN}\!-\!\phi_{i}^{SN}) \nonumber \\
&-\frac{C^{2}}{2}\sum_{j}A_{ij}\sin(\phi_{j}^{SN}\!-\!\phi_{i}^{SN})[r_{j}-r_{i}]^{2}. 
\end{align}
If we take the product of Eq.(\ref{eq:FixedPoint1p0}) with $r_{i}$, sum over $i$, and solve for $C$, we get Eq.(\ref{eq:C}) from the main text. This is easy to see by expanding $w_{i}$ in the eigenmodes of the Laplacian. In Eq.(\ref{eq:FixedPoint1p0}), the component parallel to $r_{i}$ vanishes from the saddle-node condition, while all other components vanish due to orthonormality. Note: there are two possible solutions for $C$: a stable phase-locked state (positive), and a saddle (negative). Given the definitions in the main text, the Fiedler value can be shown to be
\begin{align}
\label{eq:Fiedler}
z=-\sqrt{2R_{0}R_{2}}\kappa^{1/2},
\end{align}
by an analogous expansion. 
\subsection{\label{subsec:SCSN1} Single-cut saddle node}
The procedure outlined above can be carried out indefinitely. Here, we restrict ourselves to a particular class of saddle-node bifurcations, which we call single-cut saddle-nodes (SCSN). In this special case, a single edge becomes overloaded at bifurcation (with a $\pi/2$ phase-difference between the nodes), and is a cut edge of the network specified by the adjacency matrix, $A$. The edge partitions the network into exactly two subgraphs (denoted $\mathcal{S}_{1}$ and $\mathcal{S}_{2}$), whose nodes only share one edge (the cut edge) in common. For example, tree networks always have SCSN.

Networks with SCSN have useful properties. For instance, $K_{\text{SN}}$ can be determined by summing Eq.(\ref{eq:Swing}) over all nodes in $\mathcal{S}_{1}$ (or $\mathcal{S}_{2}$), given $\dot{v}_{i}\!=\!0$ and $v_{i}\!=\!0$ $\forall i$: 
\begin{align}
\label{eq:Ksn}
0=\sum_{i\in\mathcal{S}_{1}}P_{i} + K_{\text{SN}}\!\!\sum_{i\in\mathcal{S}_{1},j}\sin(\phi_{j}^{SN}\!-\!\phi_{i}^{SN}).
\end{align}
Only one term survives in the second sum in Eq.(\ref{eq:Ksn}): corresponding to the cut edge connecting $\mathcal{S}_{1}$ and $\mathcal{S}_{2}$-- because sine is an odd function. Moreover the phase difference for the cut edge is $\phi_{j}^{SN}\!-\!\phi_{i}^{SN}\!=\!-\pi/2$. This latter property can be shown from the saddle-node condition and the fact that for SCSN $r_{i}\!=\!\sqrt{|\mathcal{S}_{2}|/[N|\mathcal{S}_{1}|]}$ if $i\in\mathcal{S}_{1}$, and $r_{i}\!=\!-\sqrt{|\mathcal{S}_{1}|/[N|\mathcal{S}_{2}|]}$ if $i\in\mathcal{S}_{2}$. Note: the specifications for $\mathcal{S}_{1}$ and $\mathcal{S}_{2}$ are given by the convention that the node in $\mathcal{S}_{1}$ connected to $\mathcal{S}_{2}$ along the cut edge has a larger phase than its counterpart in $\mathcal{S}_{2}$. Therefore, 
\begin{align}
\label{eq:Ksn2}
K_{\text{SN}}\stackrel{\text{\tiny{SCSN}}}=\sum_{i\in\mathcal{S}_{1}}\bar{P}_{i}, 
\end{align}
as first noted for trees in \cite{Dekker}.

\begin{figure}[h]
\includegraphics[scale=0.274]{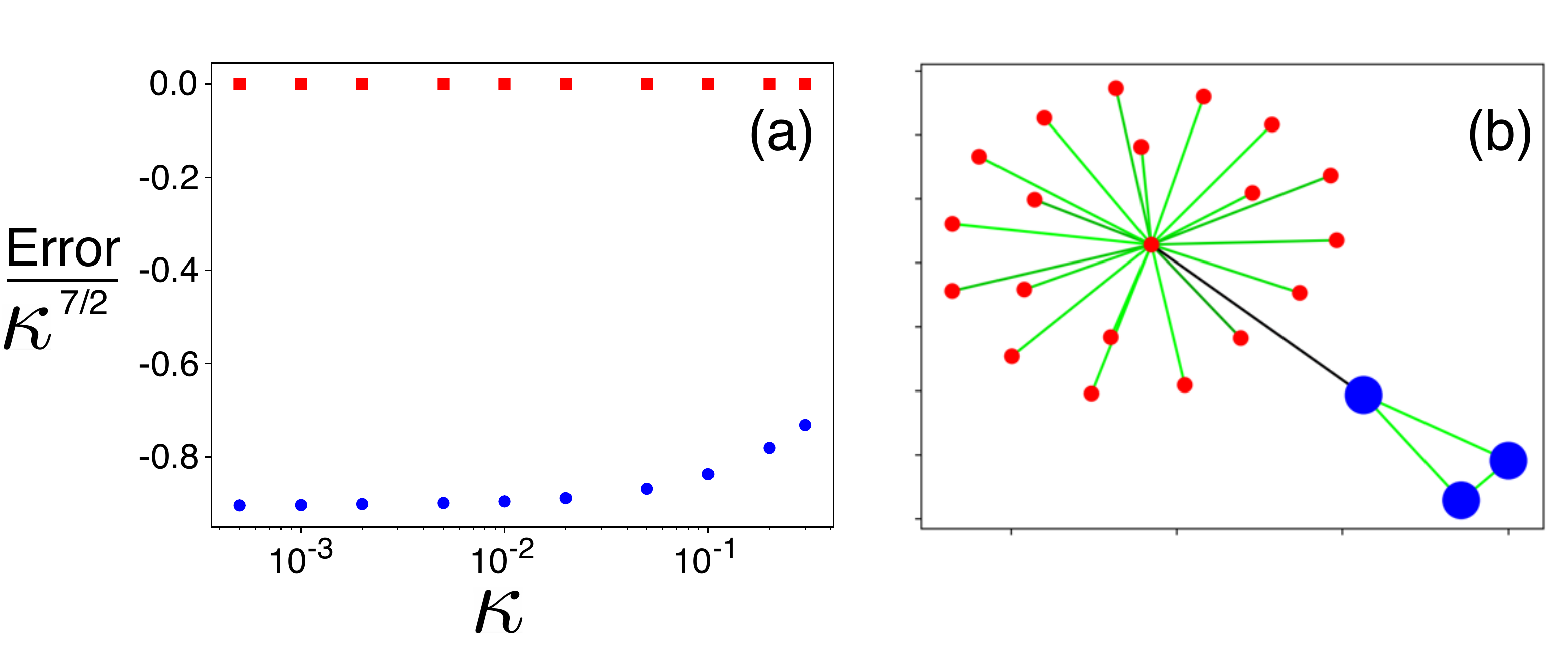}
\caption{Error scaling of Eq.(\ref{eq:FPscsn}) for a SCSN (a) mean (blue) and standard-deviation (red) of the error divided by $\kappa^{7/2}$. (b) example network.}
\label{fig:FPscaling}
\end{figure}

Using the stated SCSN properties, it is straightforward to show that
\begin{align}
\label{eq:FPscsn}
\phi_{i}^{s}-\phi_{i}^{*}= 2Cr_{i}\kappa^{1/2}\Bigg[1-\frac{5}{12}\kappa+\frac{43}{160}\kappa^{2}+...\Bigg], 
\end{align} 
by expanding the fixed-point conditions to higher orders in $\kappa$. In Fig.\ref{fig:FPscaling}(a) we plot the mean (blue) and standard deviation (red) of the error-vector, $\text{Error}_{i}\!=\![\phi_{i}^{s}-\phi_{i}^{*}-2Cr_{i}\kappa^{1/2}(1-\frac{5}{12}\kappa+\frac{43}{160}\kappa^{2})]/r_{i}$ for an example SCSN shown in Fig.\ref{fig:FPscaling}(b). We can see that the fixed-point expression Eq.(\ref{eq:FPscsn}) is accurate to $\mathcal{O}(\kappa^{7/2})$ and is parallel to $r_{i}$. The fixed points were computed numerically with the choice of zero-average-phase $\left<\phi_{i}^{s}\right>\!=\!\left<\phi_{i}^{*}\right>\!=\!0$.   
 
\section{\label{sec:SNPaths} Near bifurcation paths}
Using the fixed-point boundaries expressed in powers of $\kappa$, we can construct the optimal paths for sufficiently small $\kappa$. To make the calculation simpler, at this point we restrict ourselves to $iid$ power fluctuations, and drop the node-subscript $i$ in the noise moments, $\mu$. As mentioned in the main text,  at lowest order in $\kappa$, we substitute the ansatz $\phi_{i}(t)\!=\!\phi_{i}^{SN}\!+\!C\kappa^{1/2}r_{i}x(t)+...$ into Eqs.(\ref{eq:FullHamiltons1}-\ref{eq:FullHamiltons6}). For the other phase-space variables, we have the general expansions: 
\begin{align}
\label{eq:E1}
v_{i}&=\sum_{m}v_{i,m}\kappa^{\frac{m}{2}}, \;\;\;\;\; p_{i}=\sum_{m}p_{i,m}\kappa^{\frac{m}{2}} \nonumber \\
\lambda_{i}^{\phi}&=\sum_{m}\lambda_{i,m}^{\phi}\kappa^{\frac{m}{2}}, \;\;\;\; \lambda_{i}^{v}=\sum_{m}\lambda_{i,m}^{v}\kappa^{\frac{m}{2}}\nonumber \\
\lambda_{i}^{p}&=\sum_{m}\lambda_{i,m}^{p}\kappa^{\frac{m}{2}}, 
\end{align}
where $m\!=\!2,3,...$ When $m\!=\!2$, we assume that all variables are parallel to the Fiedler mode according to $v_{i,2}\!=\!a_{v}(x)r_{i}$, $p_{i,2}\!=\!a_{p}(x)r_{i}$, $\lambda_{i,2}^{\phi}\!=\!a_{\lambda}^{\phi}(x)r_{i}/\mu_{2}$ , $\lambda_{i,2}^{v}\!=\!a_{\lambda}^{v}(x)r_{i}/\mu_{2}$ , and $\lambda_{i,2}^{p}\!=\!a_{\lambda}^{p}(x)r_{i}/\mu_{2}$.

We note that since $\dot{x}\!=\!a_{v}(x)\kappa^{\frac{1}{2}}/C$, time-derivatives for phase-space variables other than $\phi$ are $\mathcal{O}(\kappa^{\frac{3}{2}})$, and so the LHS of Eqs.(\ref{eq:FullHamiltons2}-\ref{eq:FullHamiltons6}) can be ignored at $\mathcal{O}(\kappa)$. By substituting the expansion into Eqs.(\ref{eq:FullHamiltons2}-\ref{eq:FullHamiltons3}) and Eqs.(\ref{eq:FullHamiltons5}-\ref{eq:FullHamiltons6}), we find the following relations for $\mathcal{O}(\kappa)$:
\begin{align}
\label{eq:Delta1}
-\gamma a_{v} + a_{p} &=-K_{\text{SN}}(1-x^{2})\sum_{ij}A_{ij}\sin(\phi_{j}^{SN}\!\!-\!\phi_{i}^{SN})r_{i}\\
\label{eq:Delta2}
\alpha a_{p} &=a_{\lambda}^{p}\\
\label{eq:Delta3}
a_{\lambda}^{\phi} &=\frac{\gamma}{M} a_{\lambda}^{v}\\
\label{eq:Delta4}
\alpha a_{\lambda}^{p} &=\frac{1}{M} a_{\lambda}^{v}. 
\end{align}
One more equation is needed to close the system. We use the zero-energy condition $H(\boldsymbol{\phi},\bold{v},\bold{p},\boldsymbol{\lambda}^{\phi},\boldsymbol{\lambda}^{v},\boldsymbol{\lambda}^{p})\!=\!0$, at $\mathcal{O}(\kappa^{2})$, which results in 
\begin{align}
\label{eq:Delta5}
0=a_{\lambda}^{\phi}a_{v} -\alpha a_{\lambda}^{p}a_{p} +\frac{[a_{\lambda}^{p}]^{2}}{2}.
\end{align} 
Solving for the non-zero solution gives Eqs.(\ref{eq:OP_SN}) from the main text. Note: we have assumed that the product, $\sum_{ij}A_{ij}\sin(\phi_{j}^{SN}\!\!-\!\phi_{i}^{SN})r_{i}$, is negative in writing Eqs.(\ref{eq:OP_SN}), given our sign convention. This property appears to be general for the SN. Hence, $R_{0}\!=-\!\sum_{ij}A_{ij}\sin(\phi_{j}^{SN}\!\!-\!\phi_{i}^{SN})r_{i}\!=\!|\sum_{ij}A_{ij}\sin(\phi_{j}^{SN}\!\!-\!\phi_{i}^{SN})r_{i}|$. 
 
\subsection{\label{subsec:SCSN2}  Single-cut saddle node}
As with the fixed-point boundaries in Sec.\ref{sec:FP}, we restrict ourselves to SCSN for higher-order OPs, since the results are comparatively simple. First, we start with the ansatz $\phi_{i}(x)\!=\!\phi_{i}^{SN}\!+Cx(t)r_{i}\kappa^{1/2}[1-\frac{5}{12}\kappa]$, from Eq.(\ref{eq:FPscsn}). The next order in $\kappa$ requires substituting this ansatz, as well as the general expansion Eqs.(\ref{eq:E1}) into Eqs.(\ref{eq:FullHamiltons2}-\ref{eq:FullHamiltons3}) and Eqs.(\ref{eq:FullHamiltons5}-\ref{eq:FullHamiltons6}) and collecting terms of $\mathcal{O}(\kappa^{3/2})$. As noted in the previous section, the relevant time derivatives are at lowest order $\mathcal{O}(\kappa^{3/2})$: 
\begin{align}
\label{eq:LowestOrderDot}
\frac{\dot{v}_{i}}{r_{i}}&=\frac{\kappa^\frac{3}{2}}{C}\frac{da_{v}}{dx}a_{v}(x),\;\;\;\;\;\;\;\;\;\;\;\;  \frac{\dot{p}_{i}}{r_{i}}\!=\! \frac{\kappa^\frac{3}{2}}{C}\frac{da_{p}}{dx}a_{v}(x), \nonumber \\
\frac{\dot{\lambda}_{i}^{\phi}\mu_{2}}{r_{i}}&=\frac{\kappa^\frac{3}{2}}{C}\frac{da_{\lambda}^{\phi}}{dx}a_{v}(x), \;\;\;\;\; \frac{\dot{\lambda}_{i}^{v}\mu_{2}}{r_{i}}=\frac{\kappa^\frac{3}{2}}{C}\frac{da_{\lambda}^{v}}{dx}a_{v}(x), \nonumber \\
\frac{\dot{\lambda}_{i}^{p}\mu_{2}}{r_{i}}&=\frac{\kappa^\frac{3}{2}}{C}\frac{da_{\lambda}^{p}}{dx}a_{v}(x)
\end{align}

When $m\!=\!3$, again, all variables are parallel to the Fiedler mode according to $v_{i,3}\!=\!b_{v}(x)r_{i}$, $p_{i,3}\!=\!b_{p}(x)r_{i}$, $\lambda_{i,3}^{\phi}\!=\!b_{\lambda}^{\phi}(x)r_{i}/\mu_{2}$ , $\lambda_{i,3}^{v}\!=\!b_{\lambda}^{v}(x)r_{i}/\mu_{2}$ , and $\lambda_{i,3}^{p}\!=\!b_{\lambda}^{p}(x)r_{i}/\mu_{2}$. Substituting into Eqs.(\ref{eq:LowestOrderDot}), Eqs.(\ref{eq:FullHamiltons2}-\ref{eq:FullHamiltons3}) and Eqs.(\ref{eq:FullHamiltons5}-\ref{eq:FullHamiltons6}), we find: 
\begin{align}
\label{eq:NextDelta1}
\frac{a_{v}}{C}\frac{da_{v}}{dx}&=-\gamma b_{v} + b_{p},  \\
\label{eq:NextDelta2}
\frac{a_{v}}{C}\frac{da_{p}}{dx}&=-\alpha b_{p} + b_{\lambda}^{p}, \\
\label{eq:NextDelta3}
\frac{a_{v}}{C}\frac{da_{\lambda}^{v}}{dx}&= -b_{\lambda}^{\phi} +\frac{\gamma}{M} b_{\lambda}^{v}, \\
\label{eq:NextDelta4}
\frac{a_{v}}{C}\frac{da_{\lambda}^{p}}{dx}&= \alpha b_{\lambda}^{p} -\frac{1}{M} b_{\lambda}^{v}. 
\end{align}
Equation (\ref{eq:NextDelta1}) requires elaboration. For SCSN, the coupling term is 
\begin{align}
\label{eq:Elaboration}
&\sum_{j}\!A_{ij}\sin(\phi_{j}-\phi_{i})=\sum_{j}\!A_{ij}\sin(\phi_{j}^{SN}\!-\phi_{i}^{SN})+\nonumber \\
&C\kappa^{\frac{1}{2}}\big[1-\frac{5}{12}\kappa\big]x\sum_{j}\!A_{ij}\cos(\phi_{j}^{SN}\!-\phi_{i}^{SN})[r_{j}-r_{i}]-\nonumber \\
&\frac{C^{2}}{2!}\kappa\big[1-\frac{5}{12}\kappa\big]^{2}x^{2}\sum_{j}\!A_{ij}\sin(\phi_{j}^{SN}\!-\phi_{i}^{SN})[r_{j}-r_{i}]^{2}-\nonumber \\
&\frac{C^{3}}{3!}\kappa^{\frac{3}{2}}\big[1-\frac{5}{12}\kappa\big]^{3}x^{3}\sum_{j}\!A_{ij}\cos(\phi_{j}^{SN}\!-\phi_{i}^{SN})[r_{j}-r_{i}]^{3} +... 
\end{align}
Note: all cosine terms vanish, since if $i$ and $j$ are in the same subgraph, then $r_{i}\!=\!r_{j}$, and if they are not, then $|\phi_{j}^{SN}-\phi_{i}^{SN}|=\frac{\pi}{2}$. This means, for instance, that $\bar{P}_{i}+K_{\text{SN}}[1+\kappa]\sum_{j}A_{ij}\sin(\phi_{j}-\phi_{i})$ only contains integer powers of $\kappa$. 

To close the system, we again use the zero-energy condition at $\mathcal{O}(\kappa^{5/2})$. The resulting equation is: 
\begin{align}
\label{eq:NextDelta5}
\sum_{i}\Bigg[&\lambda_{i,3}^{\phi}v_{i,2} +\lambda_{i,2}^{\phi}v_{i,3}+\frac{1}{M}\lambda_{i,2}^{v}(-\gamma v_{i,3}+p_{i,3}) \nonumber  \\
&-\alpha p_{i,3}\lambda_{i,2}^{p}-\alpha p_{i,2}\lambda_{i,3}^{p} +\mu_{2}\lambda_{i,2}^{p}\lambda_{i,3}^{p}\Bigg].   
\end{align}
Substituting Eqs.(\ref{eq:Delta1}-\ref{eq:Delta5}) into Eqs.(\ref{eq:NextDelta1}-\ref{eq:NextDelta4}) and Eq.(\ref{eq:NextDelta5}) 
\begin{align}
\label{eq:NextDelta6}
b_{v}&=-\frac{2^{\frac{1}{2}}M}{\gamma^{3}}K_{\text{SN}}^{2}R_{0}^{\frac{3}{2}}R_{2}^{\frac{1}{2}}x[1-x^{2}]\\
\label{eq:NextDelta7}
b_{p}&=-\frac{2^{\frac{3}{2}}M}{\gamma^{2}}K_{\text{SN}}^{2}R_{0}^{\frac{3}{2}}R_{2}^{\frac{1}{2}}x[1-x^{2}]\\
b_{\lambda}^{\phi}&=0,\\
\label{eq:NextDelta8}
b_{\lambda}^{v}&=-\frac{2^{\frac{3}{2}}M^{2}\alpha^{2}}{\gamma^{2}}K_{\text{SN}}^{2}R_{0}^{\frac{3}{2}}R_{2}^{\frac{1}{2}}x[1-x^{2}],\\
\label{eq:NextDelta9}
b_{\lambda}^{p}&=-\frac{2^{\frac{3}{2}}}{\gamma}\Big[1+\frac{M\alpha}{\gamma}\Big]K_{\text{SN}}^{2}R_{0}^{\frac{3}{2}}R_{2}^{\frac{1}{2}}x[1-x^{2}]. 
\end{align}

\begin{figure}[h]
\includegraphics[scale=0.225]{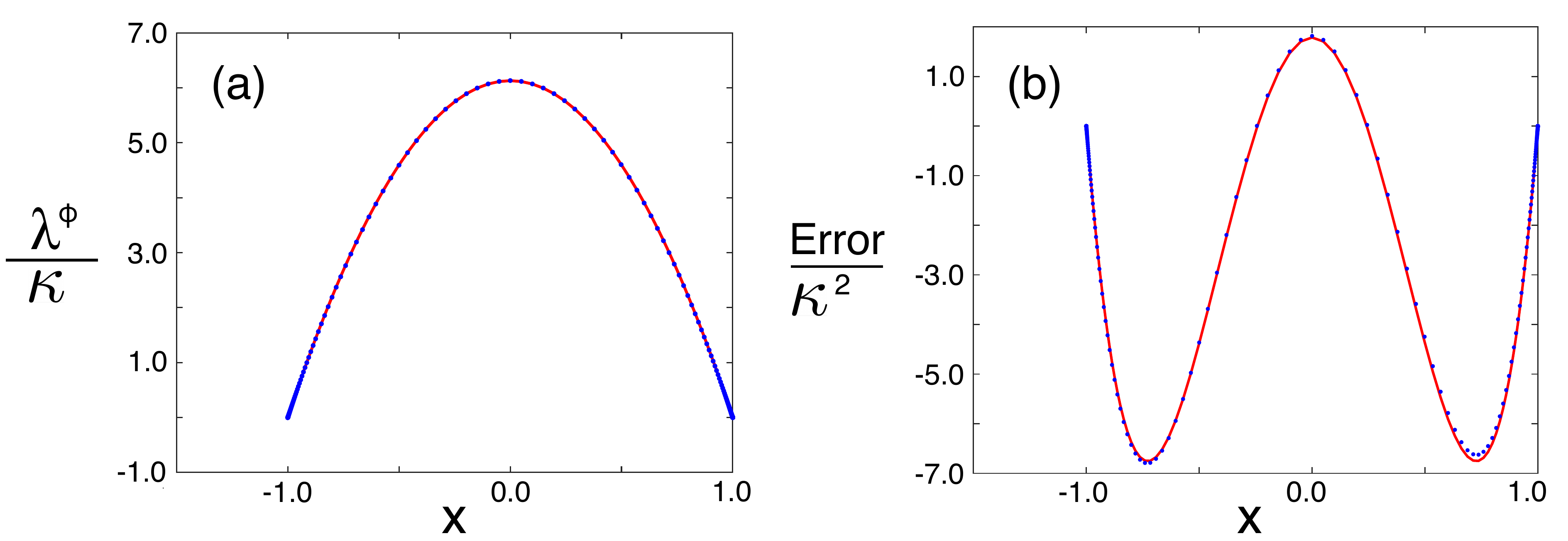}
\caption{Near bifurcation comparison for SCSN. Computed paths are shown in blue from Eqs.(\ref{eq:SCSNeff}). (a) $\mathcal{O}(\kappa)$ solution, (red) Eqs.(\ref{eq:OP_SN}). (b) $\mathcal{O}(\kappa^{2})$ solution (red) Eq.(\ref{eq:LambdaDelta2}).}
\label{fig:Lphiscaling}
\end{figure}

In order to calculate the action at next order, $\mathcal{O}(\kappa^{5/2})$, we need to know $\lambda_{i,4}^{\phi}$. This can be achieved using Eqs.(\ref{eq:NextDelta6}-\ref{eq:NextDelta9}) and the zero-energy condition at $\mathcal{O}(\kappa^{3})$. If we assume $\lambda_{i,4}^{\phi}\sim r_{i}$, then the calculation only requires a fair bit of algebra. The result is:
\begin{align}
\label{eq:LambdaDelta2}
 \frac{\lambda_{i,4}^{\phi}}{r_{i}}=& -\frac{4R_{0}^{2}\gamma\alpha^3 K_{\text{SN}}^{2}\mu_{3}[1-x^{2}]^{2}\sum_{i}r_{i}^{3}}{3\mu_{2}^{3}} \nonumber \\
  &-\frac{4R_{0}^{2}R_{2}K_{\text{SN}}^{3}x^{2}[1-x^{2}]}{\gamma\mu_{2}} \nonumber \\
   &-\frac{\gamma\alpha^{2}K_{\text{SN}}R_{2}^{\frac{1}{2}}x^{2}[1-x^{2}]}{3R_{0}^{\frac{1}{2}}\mu_{2}}. 
\end{align}
Note, that all constants in Eq.(\ref{eq:LambdaDelta2}) can be given a closed-form expression for SCSN, and depend only on $|\mathcal{S}_{1}|$ and $|\mathcal{S}_{2}|$. In Fig.\ref{fig:Lphiscaling} we show the accuracy of our expansion for the UK grid with a SCSN, Fig.\ref{fig:SSA}(b)(top).  In Fig.\ref{fig:Lphiscaling} panel (a), solutions of Eqs.(\ref{eq:SCSNeff}) are shown in blue for $\kappa\!=\!0.001$, and the $\mathcal{O}(\kappa)$ solution, Eqs.(\ref{eq:OP_SN}), is shown in red. In panel (b), we plot the difference between the two curves in panel (a) in blue, and compare to the $\mathcal{O}(\kappa^{2})$ solution Eq.(\ref{eq:LambdaDelta2}) shown in red. The agreement is excellent at both orders of $\kappa$. 

Integrating our expressions for the phase-space variables to $\mathcal{O}(\kappa^{2})$ in $\lambda_{i}^{\phi}$ and $\mathcal{O}(\kappa^{\frac{3}{2}})$ for all others, according to Eq.(\ref{eq:Action}), gives the action for SCSN at $\mathcal{O}(\kappa^{\frac{5}{2}})$.  

\section{\label{sec:NGE}  Non-Gaussian effects}
We would like to know how the higher-moments of the noise distribution contribute to the action in general.  As in the main text, we restrict ourselves to $iid$ power fluctuations. A clue comes from our solution for the OP at lowest order in $\kappa$. Recall from Sec.\ref{sec:SNPaths}, that $\lambda_{i,2}^{\phi}$ is determined by the zero-energy condition at $\mathcal{O}(\kappa^{2})$ -- where the noise variance $\mu_{2}$ first appears in $H(\boldsymbol{\phi},\bold{v},\bold{p},\boldsymbol{\lambda}^{\phi},\boldsymbol{\lambda}^{v},\boldsymbol{\lambda}^{p})$. The general pattern is the following: the contribution proportional to $\mu_{n}$ in $\lambda_{i,2(n-1)}^{\phi}$ is determined by the zero-energy condition at $\mathcal{O}(\kappa^{n})$. First, we can check that this pattern holds for $\mu_{3}$, by looking at $H(\boldsymbol{\phi},\bold{v},\bold{p},\boldsymbol{\lambda}^{\phi},\boldsymbol{\lambda}^{v},\boldsymbol{\lambda}^{p})\!=\!0$ at  $\mathcal{O}(\kappa^{3})$. Using our general expansion, we get an equation:
\begin{align}
\label{eq:ThirdOrder}
0=&\sum_{i}\Bigg[\lambda_{i,4}^{\phi}v_{i,2}+\lambda_{i,3}^{\phi}v_{i,3}+\lambda_{i,2}^{\phi}v_{i,4} +\frac{1}{M}\lambda_{i,4}^{v}T_{i,2}+\nonumber \\
&\frac{1}{M}\lambda_{i,3}^{v}T_{i,3}+\frac{1}{M}\lambda_{i,2}^{v}T_{i,4}-\alpha\lambda_{i,4}^{p}p_{i,2}-\alpha\lambda_{i,3}^{p}p_{i,3}-\nonumber \\
&\alpha\lambda_{i,2}^{p}p_{i,4}+\mu_{2}\lambda_{i,2}^{p}\lambda_{i,4}^{p}+\frac{\mu_{2}}{2}[\lambda_{i,3}^{p}]^{2}+\frac{\mu_{3}}{3!}[\lambda_{i,2}^{p}]^{3}\Bigg].
 \end{align}
 Note, we have introduced the notation $T_{i}\!\equiv\!-\gamma v_{i}+p_{i}+\bar{P}_{i}+K\sum_{j}A_{ij}\sin(\phi_{j}-\phi_{i})\!$ and $T_{i}\!=\!\sum_{m}T_{i,m}\kappa^{\frac{m}{2}}$. 
 
It is important to realize a few properties of the phase-space coordinates. First, only the conjugate momenta, (the $\lambda$'s), depend explicitly on the noise statistics. Second, since $\mathcal{O}(\kappa^{n})$ is the lowest order at which $\mu_{n}$ enters the Hamiltonian, $\lambda_{i,2(n-1)-j}^{\phi}$, $\lambda_{i,2(n-1)-j}^{v}$, and $\lambda_{i,2(n-1)-j}^{p}$ $\forall j \in\{1,2,...,2(n-2)\}$ depend on $\mu_{n-1}$, $\mu_{n-2}$,... $\mu_{2}$, but not on $\mu_{n}$. Therefore, if $\lambda_{i,4}^{v}$ and $\lambda_{i,4}^{p}$ vanish from Eq.(\ref{eq:ThirdOrder}), then only two terms depend on $\mu_{3}$, i.e., $\lambda_{i,4}^{\phi}v_{i,2}$ and $\frac{\mu_{3}}{3!}[\lambda_{i,2}^{p}]^{3}$, and we can solve explicitly for the contribution to $\lambda_{i,4}^{\phi}$ from $\mu_{3}$. 
 
Luckily, the terms proportional to $\lambda_{i,4}^{v}$ and $\lambda_{i,4}^{p}$ in Eq.(\ref{eq:ThirdOrder}) depend on the lowest order solution, which we have explicit expressions for. In fact, $T_{i,2}\!=\!0$ and $-\alpha p_{i,2}+\mu_{2}\lambda_{i,2}^{p}\!=\!0$. Finally, we add an additional assumption that $\lambda_{i,4}^{\phi}\sim r_{i}$, as with the SCSN, Eq.(\ref{eq:LambdaDelta2}). With this assumption, the lowest-order (in $\kappa$) contribution to $\lambda_{i}^{\phi}$ from $\mu_{3}$, denoted $\Delta^{(3)}\lambda_{i}^{\phi}$ is found from Eq.(\ref{eq:ThirdOrder}): 
\begin{align}
\label{eq:LambdaThird}
\frac{\Delta^{(3)}\lambda_{i}^{\phi}}{r_{i}}=-\frac{\mu_{3}\sum_{j}[\lambda_{j,2}^{p}]^{3}}{3!\sum_{j}r_{j}v_{j,2}}.
\end{align}
Exactly the same argument can be used to calculate $\Delta^{(4)}\lambda_{i}^{\phi}$, etc. The general expression is 
\begin{align}
\label{eq:LambdaThird}
\frac{\Delta^{(n)}\lambda_{i}^{\phi}}{r_{i}}=-\frac{\mu_{n}\sum_{j}[\lambda_{j,2}^{p}]^{n}}{n!\sum_{j}r_{j}v_{j,2}}. 
\end{align}
 
Finally, the lowest-order contribution to the action from $\mu_{n}$ is found from Eq.(\ref{eq:Action}) -- namely, integrating $\Delta^{(n)}\lambda_{i}^{\phi}(x)$ over $x$. The result is Eq.(\ref{eq:Correction}) in the main text. Note: the integral contributions from $\lambda_{i}^{v}$ and $\lambda_{i}^{p}$ can be ignored, since $v_{i}$ and $p_{i}$ are at lowest order $\mathcal{O}(\kappa)$, while $\phi_{i}^{s}-\phi_{i}^{*}$ is $\mathcal{O}(\kappa^{1/2})$. 

Furthermore, Eq. (\ref{eq:Correction}) can be used to compare the actions of two increment distributions that differ in the $n$th moment. For instance, the difference in the action between a symmetric increment distribution ($\mu_{3}\!=\!0$) and a non-symmetric distribution ($\mu_{3}\!\neq\!0$), each with the same variance $\mu_{2}$, is given by Eq.(\ref{eq:Correction}) with $n=3$. Moreover, as noted in the main text, if the power fluctuations are assumed to be Gaussian white noise with a variance $\mu_{2}$, then Eq. (\ref{eq:Correction}) gives the exponential correction to the rate of desynchronization for non-symmetric $(n=3)$, and symmetric $(n=4)$ increment distributions. Our expansion gives very accurate results for desynchronization rates near bifurcation, as demonstrated in Fig.\ref{fig:StatMatch}. 

For completeness, we note that for Gaussian white noise, Hamilton's equations are identical, except for Eq.(\ref{eq:FullHamiltons3}):
\begin{align}
\label{eq:GaussianPower}
\dot{p}_{i}\stackrel{\text{\tiny{Guass}}}=-\alpha p_{i} +\mu_{i,2}\lambda_{i}^{p},
\end{align} 
and the Hamiltonian Eq.(\ref{eq:Hamiltonian}):
\begin{align}
&H(\boldsymbol{\phi},\bold{v},\bold{p},\boldsymbol{\lambda}^{\phi},\boldsymbol{\lambda}^{v},\boldsymbol{\lambda}^{p})\!\stackrel{\text{\tiny{Guass}}}=\sum_{i}\!\!\Bigg[\;\lambda^{\phi}_{i}v_{i}-\alpha p_{i}\lambda^{p}_{i}+\frac{\mu_{i,2}}{2}[\lambda_{i}^{p}]^{2}\nonumber \\ 
&+\frac{\lambda_{i}^{v}}{M}\Big(\!\! -\gamma v_{i} + \bar{P}_{i}+p_{i}+\!\sum_{j}K_{ij}\sin(\phi_{j}-\phi_{i}) \!\Big)\Bigg].
\label{eq:HamiltonianGauss}  
\end{align}
\section{\label{sec:SSA} Synchronized subgraph approximation}
As mentioned in the main text, in many cases the SN bifurcation occurs with exact symmetry in the Fiedler mode; namely, the network is partitioned into $\mathcal{N}$ subgraphs, $\mathcal{S}_{1}, \mathcal{S}_{2},...,\mathcal{S}_{\mathcal{N}}$, where two nodes $i$ and $j$ in the same subgraph $\mathcal{S}_{n}$ have $r_{i}\!=\!r_{j}=r^{(n)}$ at bifurcation. As already noted, SCSN cases have $\mathcal{N}=2$, including trees. Larger values of $\mathcal{N}$ occur for block-networks (or clique trees), such as Fig.\ref{fig:SSA}(b)--the lower panel.

Since, the Fiedler mode is the weakest stable mode of the network and the optimal desynchronization path is parallel to $r_{i}$ over several orders in $\kappa$, as demonstrated in previous sections, we simply assume (as an approximation) that all nodes within the same subgraph, e.g., $n$, are synchronized: $\phi_{i}-\phi_{i}^{SN}=\Phi_{n}$, $v_{i}=\mathcal{V}_{n}$, $p_{i}=\mathcal{P}_{n}$, $\lambda_{i}^{\phi}=l_{n}^{\phi}$, $\lambda_{i}^{v}=l_{n}^{v}$, and $\lambda_{i}^{p}=l_{n}^{p}$ $\forall i \in\mathcal{S}_{n}$. We can find an approximate set of Hamilton's equations for such OPs by simply averaging over all nodes within a subgraph: 
\begin{align}
\label{eq:Average}
\dot{\Phi}_{n}&=\sum_{i\in\mathcal{S}_{n}}\frac{\dot{\phi}_{i}}{|\mathcal{S}_{n}|}, \;\;\; \dot{\mathcal{V}}_{n}=\sum_{i\in\mathcal{S}_{n}}\frac{\dot{v}_{i}}{|\mathcal{S}_{n}|},\nonumber \\
\dot{\mathcal{P}}_{n}&=\sum_{i\in\mathcal{S}_{n}}\frac{\dot{p}_{i}}{|\mathcal{S}_{n}|}, \;\;\;\; \dot{l}_{n}^{\phi}=\sum_{i\in\mathcal{S}_{n}}\frac{\dot{\lambda}_{i}^{\phi}}{|\mathcal{S}_{n}|} \nonumber \\
\dot{l}_{n}^{v}&=\sum_{i\in\mathcal{S}_{n}}\frac{\dot{\lambda}_{i}^{v}}{|\mathcal{S}_{n}|} \;\;\;\;\;\; \dot{l}_{n}^{p}=\sum_{i\in\mathcal{S}_{n}}\frac{\dot{\lambda}_{i}^{p}}{|\mathcal{S}_{n}|}.
\end{align}

Performing these averages over the original Eqs.(\ref{eq:FullHamiltons1}-\ref{eq:FullHamiltons6}) results in network coupling terms 
\begin{align}
\label{eq:Coupling1}
\mathcal{X}_{n}&=\sum_{i\in\mathcal{S}_{n}, j\notin\mathcal{S}_{n}}A_{ij}\sin(\phi_{j}-\phi_{i})\;\;\; \text{and}\\
\label{eq:Coupling2}
\mathcal{Y}_{n}&=\sum_{i\in\mathcal{S}_{n}, j\notin\mathcal{S}_{n}}A_{ij}\cos(\phi_{j}-\phi_{i})[\lambda_{j}^{v}-\lambda_{i}^{v}]
\end{align}
Let us specify an index function which maps the node number $i$ to its subgraph number $n$, i.e., $F(i)\!=\!n$. Using this notation, the coupling terms become:  
\begin{align}
\label{eq:Coupling3}
\mathcal{X}_{n}=\!\!\!\!\sum_{i\in\mathcal{S}_{n}, j\notin\mathcal{S}_{n}}\!\!\!\!A_{ij}&\Big[\sin(\phi_{j}^{SN}\!\!-\!\phi_{i}^{SN})\cos(\Phi_{F(j)}-\Phi_{F(i)}) \nonumber \\ 
&+\cos(\phi_{j}^{SN}-\phi_{i}^{SN})\sin(\Phi_{F(j)}-\Phi_{F(i)})\Big] 
\end{align}
\begin{align}
\label{eq:Coupling4}
\mathcal{Y}_{n}=\!\!\!\!&\sum_{i\in\mathcal{S}_{n}, j\notin\mathcal{S}_{n}}\!\!\!\!\!\!\!\!A_{ij}\!\Big[\!\cos(\phi_{j}^{SN}\!\!\!-\!\phi_{i}^{SN})\!\sin(\Phi_{F(j)}\!-\!\Phi_{F(i)})[l^{v}_{F(j)}\!-\!l^{v}_{F(i)}] \nonumber \\ 
&-\sin(\phi_{j}^{SN}\!\!\!-\!\phi_{i}^{SN})\!\cos(\Phi_{F(j)}\!-\!\Phi_{F(i)})[l^{v}_{F(j)}\!-\!l^{v}_{F(i)}]\Big]. 
\end{align}

Now, since all phase-space variables are assumed to be synchronized within subgraphs, we can define the following coupling matrices between subgraphs $n$ and $n'$
\begin{align}
\label{eq:Coupling5}
S_{nn'}&=\sum_{i\in\mathcal{S}_{n}, j\in\mathcal{S}_{n'}}\!\!\!\!\!A_{ij}\sin(\phi_{j}^{SN}-\phi_{i}^{SN}),\\
\label{eq:Coupling6}
C_{nn'}&=\sum_{i\in\mathcal{S}_{n}, j\in\mathcal{S}_{n'}}\!\!\!\!\!A_{ij}\cos(\phi_{j}^{SN}-\phi_{i}^{SN}).
\end{align}
Therefore, the synchronized subgraph equations become
\begin{align}
\label{eq:SSA1}
\dot{\Phi}_{n}&=\mathcal{V}_{n}, \\
\label{eq:SSA2}
M\dot{\mathcal{V}}_{n}&= -\gamma\mathcal{V}_{n}+\mathcal{P}_{n} \!+\!\sum_{i\in\mathcal{S}_{n}}\!\frac{\bar{P}_{i}}{{|\mathcal{S}_{n}|}} \;\;+\nonumber \\
&\!\!\!\!\!\!\!\!\!\frac{K_{\text{SN}}[1+\kappa]}{|\mathcal{S}_{n}|}\!\!\sum_{n'\neq n}\!S_{n n'}\cos(\Phi_{n'}\!-\!\Phi_{n})\!+\!C_{n n'}\sin(\Phi_{n'}\!-\!\Phi_{n}), \\
\label{eq:SSA3}
\dot{\mathcal{P}}_{n}&=-\alpha\mathcal{P}_{n} +\sum_{m}\nu_{m}g_{m}\exp \{ g_{m}l_{n}^{p} \} \\
\label{eq:SSA4}
\dot{l}_{n}^{\phi}&=-\frac{K_{\text{SN}}[1+\kappa]}{M|\mathcal{S}_{n}|}\!\!\sum_{n'\neq n}\!\Bigg[C_{n n'}\cos(\Phi_{n'}\!-\!\Phi_{n})[l_{n'}^{v}-l_{n}^{v}] \nonumber \\
&\;\;\;\;\;\;\;\;\;\;\;\;\;\;\;\;\;\;\;\;\;\;\;\;\;-S_{n n'}\sin(\Phi_{n'}\!-\!\Phi_{n})[l_{n'}^{v}-l_{n}^{v}] \Bigg],\\
\label{eq:SSA5}
\dot{l}_{n}^{v}&=-l_{n}^{\phi} +\frac{\gamma}{M}l_{n}^{v},\\
\label{eq:SSA6}
\dot{l}_{n}^{p}&=\alpha l_{n}^{p} -l_{n}^{v}/M.  
\end{align} 
In general, the matrices Eqs.(\ref{eq:Coupling5}-\ref{eq:Coupling6}) have to be computed numerically at the saddle-node bifurcation in order to solve Eqs.(\ref{eq:SSA1}-\ref{eq:SSA6}). 

However, for SCSN the results simplify significantly. In this case $n\!=\!1,2$, and $C_{2,1}=C_{1,2}=0$ and  $S_{2,1}=-S_{1,2}=1$. In addition, we notice that the effective two-oscillator system for SCSN can be reduced to one by introducing the relative coordinates: $\Psi\!=\!\Phi_{1}\!-\!\Phi_{2}$, $V\!=\!\dot{\Psi}$, $P\!=\!\mathcal{P}_{1}\!-\!\mathcal{P}_{2}$, $\Lambda_{\Psi}\!=\!|\mathcal{S}_{1}|l_{1}^{\phi}\!=\!-|\mathcal{S}_{2}|l_{2}^{\phi}$, $\Lambda_{V}\!=\!|\mathcal{S}_{1}|l_{1}^{v}\!=\!-|\mathcal{S}_{2}|l_{2}^{v}$, and $\Lambda_{P}\!=\!|\mathcal{S}_{1}|l_{1}^{p}\!=\!-|\mathcal{S}_{2}|l_{2}^{p}$. The result is Eqs.(\ref{eq:SCSNeff}) in the main text. 

\begin{figure}[t]
\includegraphics[scale=0.40]{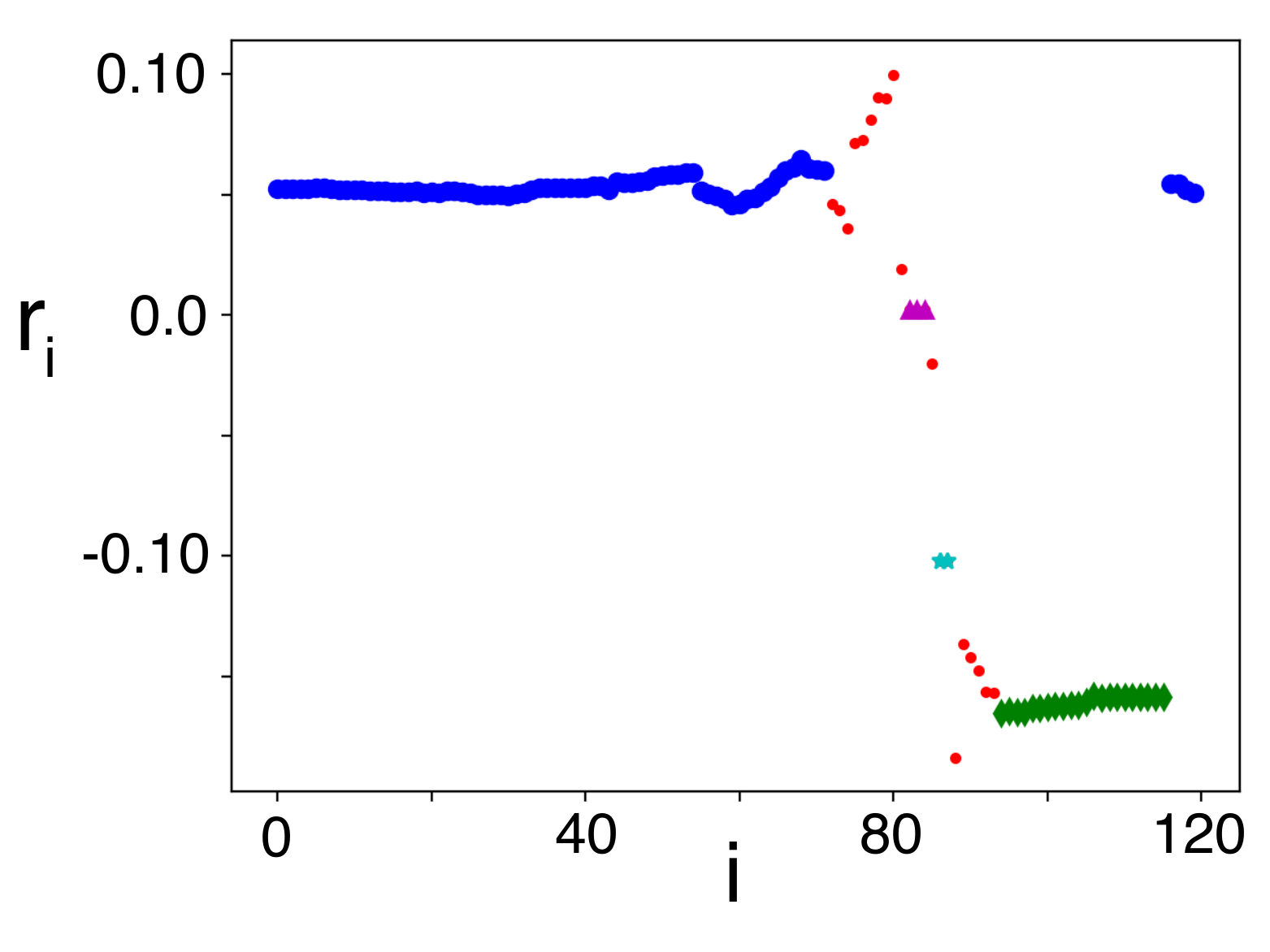}
\caption{Fiedler mode at bifurcation for the UK grid-example from Fig.\ref{fig:SSA} without symmetry (magenta). Blue, magenta, green, and cyan nodes were placed in the same subgraphs, since their values are within $5\%$. Each of the red nodes was treated as a subgraph.}
\label{fig:Fiedler}
\end{figure}

In addition, Eq.(\ref{eq:SSA1}-\ref{eq:SSA6}) can be used as a coarse-grained approximation for networks without exact symmetry in the Fiedler mode at bifurcation. One simply groups together nodes in a subgraph if they have similar Fiedler-mode values, where similar means within some tolerance $\epsilon$, e.g., $\big|\!\sum_{l\in\mathcal{S}_{n}}\!r_{l}/|\mathcal{S}_{n}|\!-\!r_{i}\big|\!<\!\epsilon$. An example is shown in Fig.\ref{fig:Fiedler}, for the UK power-grid example from Fig.\ref{fig:SSA} without symmetry at bifurcation (magenta). Plotted is the Fiedler mode at bifurcation versus the node number, $i$. Blue, magenta, green, and cyan nodes were placed in the same subgraphs, since node Fiedler-mode values were within $5\%$ of one another (for similarly colored nodes). Each of the red nodes was treated as a subgraph (of size one). The total number of subgraphs was twenty given this partition. The coupling matrices Eqs.(\ref{eq:Coupling5}-\ref{eq:Coupling6}) were computed at bifurcation, and the OPs were solved from Eqs.(\ref{eq:SSA1}-\ref{eq:SSA6}) with twenty subgraphs.

 \section{\label{sec:Hetero} Parameter heterogeneity}
In this section we show that our expansion in $\kappa$ and the SSA are robust to parameter variation. Instead of being homogeneous, we let $M$, $\gamma$, and $\alpha$ be drawn from uniform distributions, independently for each node. Given a heterogeneity scale parameter $\eta$, we let $M_{i}\!=\!M_{0}[1.0+\eta(\mathcal{R}_{i,M}-0.5)]$, $\gamma_{i}\!=\!\gamma_{0}[1.0+\eta(\mathcal{R}_{i,\gamma}-0.5)]$, and $\alpha_{i}\!=\!\alpha_{0}[1.0+\eta(\mathcal{R}_{i,\alpha}-0.5)]$, where $\mathcal{R}_{i,M}$, $\mathcal{R}_{i,\gamma}$, and $\mathcal{R}_{i,\alpha}$ are independent, random numbers drawn uniformly over $[0,1]$. In this section , we take $M_{0}\!=\!0.02546s^{2}$, $\gamma_{0}\!=\!0.10053s$, and $\alpha_{0}\!=\!1s^{-1}$.
 
Figure \ref{fig:Inhom} compares predictions of Eq.(\ref{eq:Correction}) with the computed difference in the actions between Gaussian and non-Gaussian noise. The computed actions were found by solving Eqs.(\ref{eq:FullHamiltons1}-\ref{eq:FullHamiltons6}). Note: in Eqs.(\ref{eq:FullHamiltons1}-\ref{eq:FullHamiltons6}) $M\!\rightarrow\!M_{i}$, $\alpha\!\rightarrow\!\alpha_{i}$ and $\gamma\!\rightarrow\!\gamma_{i}$. Results are shown for several levels of the heterogeneity parameter $\eta$ and are labeled by the average coefficient of variation, i.e, $\text{cv}\!\equiv\!\big[\frac{\sigma_{M}}{\left<M\right>}+\frac{\sigma{\gamma}}{\left<\gamma\right>}+\frac{\sigma_{\alpha}}{\left<\alpha\right>}\big]/3$, where $\sigma_{q}$ and $\left<q\right>$ denote the standard deviation and average for parameter $q$. The underlying network corresponds to Fig.\ref{fig:StatMatch}(b). We see good agreement between numerics and Eq.(\ref{eq:Correction}) despite significant heterogeneity, particularly for skewed noise distributions.
\vspace{0.3cm}
\begin{figure}[t]
\includegraphics[scale=0.43]{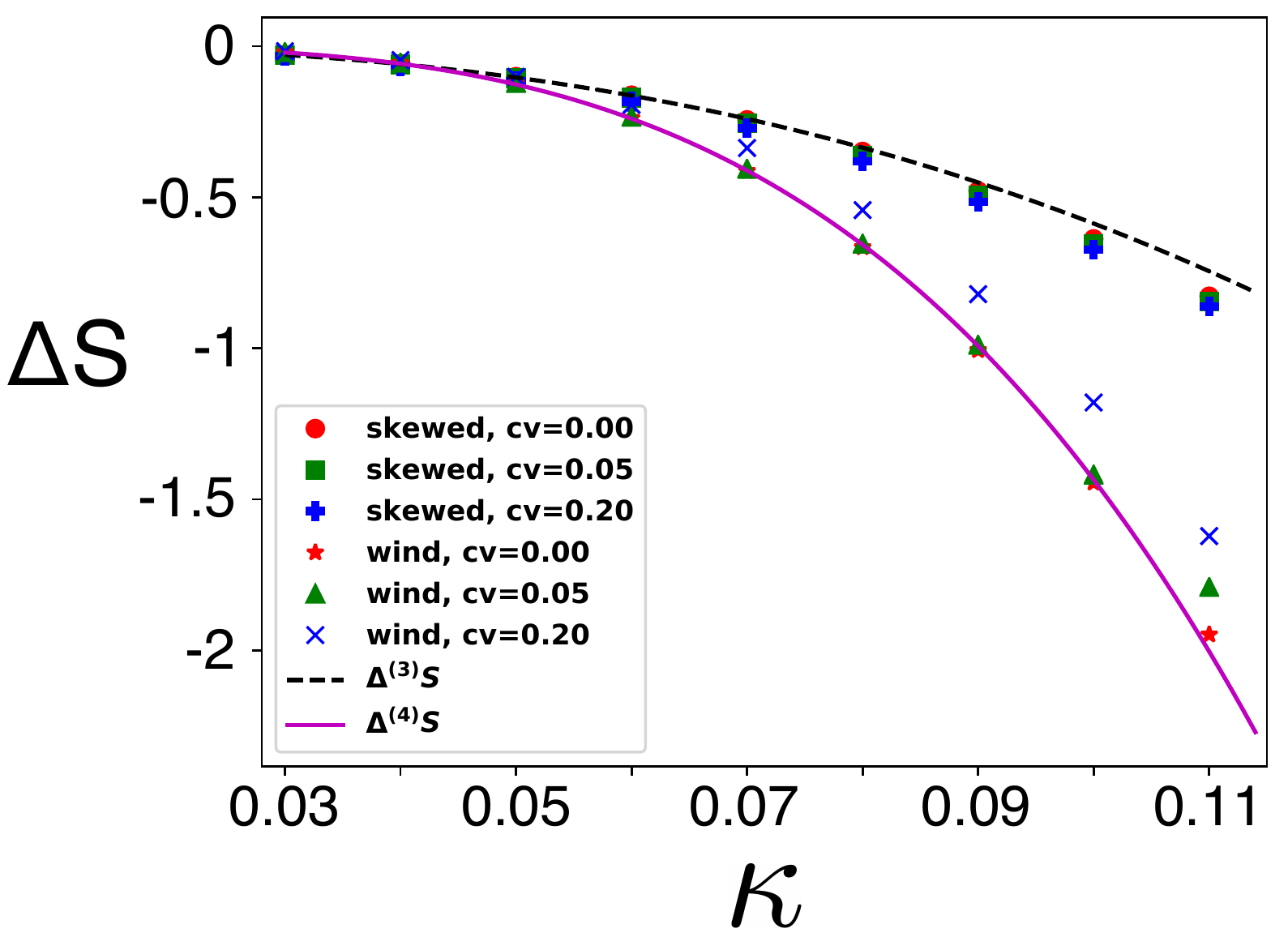}
\caption{Difference in the action between Gaussian and non-Gaussian noise for several values of parameter heterogeneity.}
\label{fig:Inhom}
\end{figure}     
 
Similarly, we perform computations using the SSA for single-cut-saddle-nodes with homogeneous parameters, Eqs.(\ref{eq:SCSNeff}-\ref{eq:ActionSCSN}), and compare with heterogeneous parameters. For example, Figure \ref{fig:Inhom} shows such a compare for the network drawn in Fig.\ref{fig:FPscaling}(b). The computed actions are shown in blue and the SSA results are shown with a black-line for skewed noise (see Sec.\ref{sec:Sims}). Again, we see good agreement despite the heterogeneity.
\begin{figure}[h]
\vspace{0.7cm}
\includegraphics[scale=0.43]{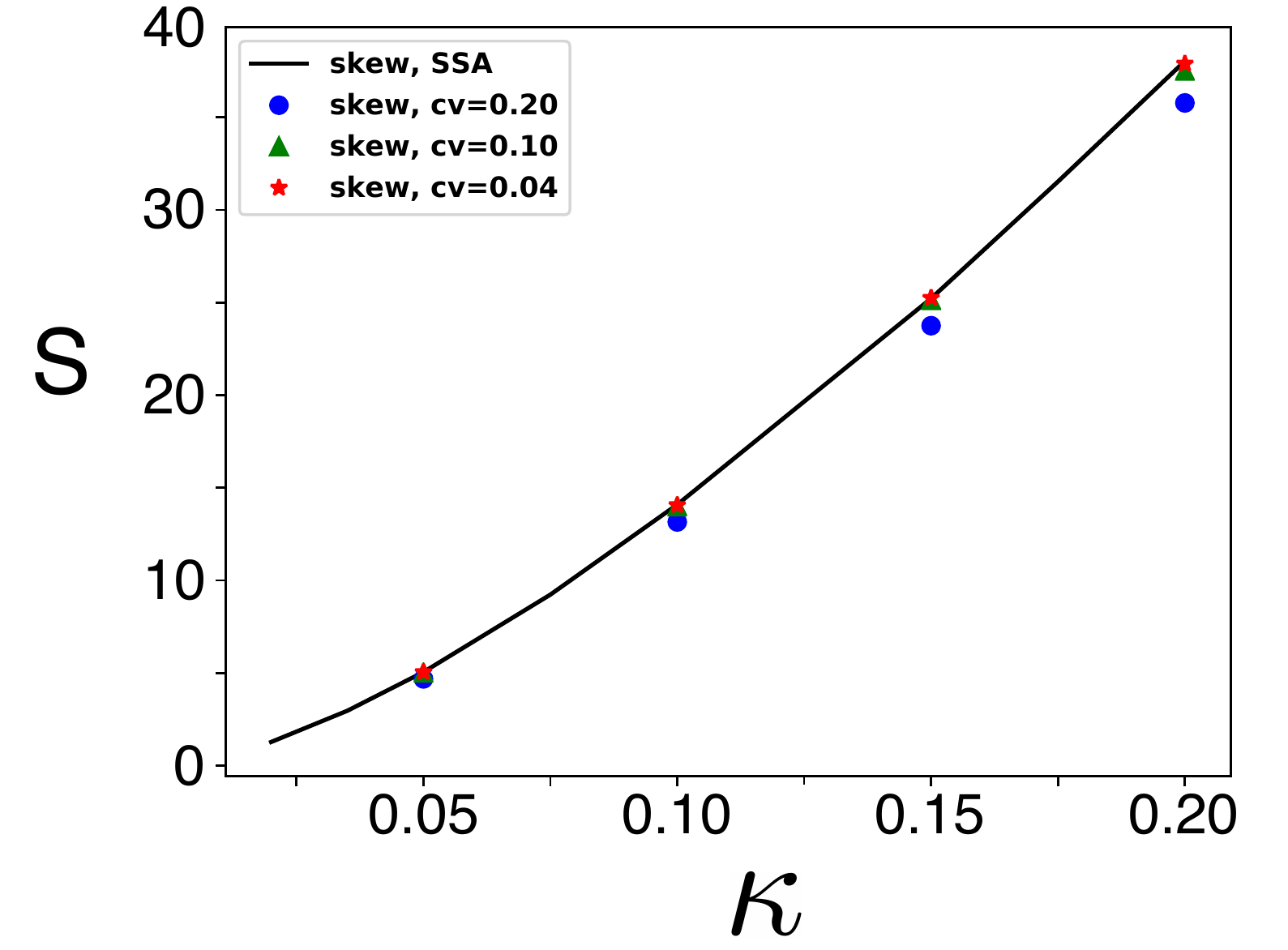}
\caption{Action for a network with a single over-loaded edge, i.e., Fig.\ref{fig:FPscaling}(b), with parameter heterogeneity.}
\label{fig:Inhom}
\end{figure} 

\section{\label{sec:Sims}Simulations} 
Monte-Carlo simulations were performed using a combination of Gillespie's algorithm for power fluctuations, $p_{i}$, and Euler's method for integrating Eqs.(\ref{eq:Swing}-\ref{eq:Pulse}) between reaction times. For illustration, let us assume that the noise is independent and identically distributed (iid) for each node. The stochastic rate for the {\it next reaction} is $\mathcal{R}\!=\!\sum_{i,m}\nu_{im}=N\sum_{m}\nu_{m}$. 
The next reaction time is stochastically selected $\Delta T\!=-\!\ln(r_{1})/\mathcal{R},$ where $r_{1}$ is a uniformly distributed random number over the unit interval. Since the noise is $iid$, the node which receives the increment is selected uniformly at random from the $N$ nodes. Another random number is generated, $\mathcal{P}\!=\!r_{2}\sum_{m}\nu_{m}$, where $r_{2}$ is a uniformly distributed random number over the unit interval. The $n$th increment is chosen if $\sum_{m=1}^{n-1}\nu_{m} < \mathcal{P} < \sum_{m=1}^{n}\nu_{m}$. Equations (\ref{eq:Swing}-\ref{eq:Pulse}) are integrated with Euler's method from $t$ to $t+\Delta T$ with time steps $dt\!=\!4*10^{-5}$, at which time the selected node, e.g,. $i$, has its power incremented: $p_{i}\!\rightarrow\!p_{i}\!+\!g_{n}$. On the other hand, for Gaussian noise the Euler-Milstein method was used with $dt\!=\!4*10^{-5}$.  

The wind-turbine data was taken from \cite{Haehne}. The data consists of power measurements at 1-second resolution for 12 turbines, each rated at 2MW. We averaged over the available data at each time step and histogrammed the power increments $p(t+1)\!-\!p(t)$ using $60$ uniformly spaced bins ($\mathcal{M}\!=\!60$). The result is shown Fig.\ref{fig:Noise}(a). In order to make desynchronization less rare, we artificially increased the occurrence rate for the Poisson pulses, such that $\sum_{m}\nu_{m}\!=\!10s^{-1}$ (not $1s^{-1}$). For reference, the noise variance for $iid$ wind-turbine noise given our model is $\mu_{2}\!=\!5.8317*10^{-4}s^{-1}$. On the other hand for skewed noise comparisons, we chose a simple two-pulse model: $\nu_{m}\!\in\!\{10,3.33333\}$ and $g_{m}\!\in\!\{-0.0038183,0.0114549\}$, which has the same variance as the wind-turbine model.     

Lastly, each Monte-Carlo point in Figs.\ref{fig:StatMatch}-\ref{fig:SSA} represent the log of the average of 200 slip times -- defined as the time it takes to see a phase difference greater than $2\pi$ develop between any connected oscillators. Each of the slips were generated from the phase-locked state initial conditions and with different random number seeds. All simulations in Fig.\ref{fig:SSA} were done with skewed noise except for the block network (cyan), for which we used the wind-turbine noise.








\begin{thebibliography}{9}
\bibitem{FN5} Equation \ref{eq:Swing} is dimensionless-- normalized by rated power. The inertia and damping constants are fixed\cite{Timme2}: $M\!=\!0.02546s^{2}$ and $\gamma\!=\!0.10053s$. The average powers, $\bar{P}_{i}$, are independent, identically, and uniformly distributed over the interval $[-0.5,0.5]$. The power-damping rate, $\alpha$, was taken to range from $0.1s^{-1}$ to $1s^{-1}$. 
\bibitem{FNtime} The time constant for the wind-turbine fluctuations was chosen to be a factor of ten smaller than the original data, i.e., $\tau\!=\!0.1s$ , so that a wider range of parameters could be explored.
\bibitem{FN2} The energy along the optimal path is zero since there is no explicit time dependence in Eqs.(\ref{eq:Swing}-\ref{eq:Pulse}).
\end{thebibliography}
\end{document}